\documentclass[11pt,peerreviewca,draftclsnofoot]{IEEEtran}%
\usepackage{here}
\usepackage{amsmath}
\usepackage{amsfonts}
\usepackage{amssymb}
\usepackage{graphicx}
\usepackage{color}
\usepackage{fancybox}
\usepackage{url}%
\providecommand{\U}[1]{\protect\rule{.1in}{.1in}}
\newtheorem{theorem}{Theorem}

\newtheorem{corollary}[theorem]{Corollary}

\newtheorem{definition}[theorem]{Definition}

\newtheorem{proposition}[theorem]{Proposition}

\begin{document}

\title{A Multi-Scan Labeled Random Finite Set Model for Multi-object State Estimation}
\author{Ba-Tuong~Vo, Ba-Ngu~Vo \thanks{Acknowledgement: This work is supported by the
Australian Research Council under Discovery Projects DP170104854 and
DP160104662.}}
\maketitle

\begin{abstract}
State space models in which the system state is a finite set--called the
multi-object state--have generated considerable interest in recent years.
Smoothing for state space models provides better estimation performance than
filtering by using the full posterior rather than the filtering density. In
multi-object state estimation, the Bayes multi-object filtering recursion
admits an analytic solution known as the Generalized Labeled Multi-Bernoulli
(GLMB) filter. In this work, we extend the analytic GLMB recursion to
propagate the multi-object posterior. We also propose an implementation of
this so-called multi-scan GLMB posterior recursion using a similar approach to
the GLMB filter implementation.

\end{abstract}

%\markboth{Preprint: IEEE Transactions on Signal Processing,~Vol.~60, No.~1, pp. 2--17, ~JAN~2012}{Vo et al.: Closed form solutions to %forward-backward smoothing}

\section{Introduction$\label{sec:intro}$}

In Bayesian estimation for state-space models, smoothing yields significantly
better estimates than filtering by using the history of the states rather than
the most recent state \cite{Meditch73}, \cite{Briers08}, \cite{Doucetoverview}%
. Conditional on the observation history, filtering only considers the current
state via the \emph{filtering density}, whereas smoothing considers the
sequence of states up to the current time via the \emph{posterior density}.
Numerical methods for computing the filtering and posterior densities have a
long history and is still an active area of research, see for example
\cite{AndersonMoore79}, \cite{Ristic_etal_04}, \cite{DoucetGodsillAndrieu00},
\cite{Doucetoverview}. Recursive computation of the posterior density is also
known as \emph{smoothing-while-filtering} \cite{Briers08}.

A generalisation of state-space models that has attracted substantial interest
in recent years is Mahler's Finite Set Statistics (FISST) framework for
multi-object system \cite{MahlerPHD}, \cite{MahlerCPHD}, \cite{Mahler07,
Mahler14}. Instead of a vector, the state of a multi-object system at each
time, called the \emph{multi-object state}, is a finite set of vectors. Since
its inception, a host of algorithms have been developed for \emph{multi-object
state estimation }\cite{Mahler07, Mahler14}. By incorporating labels (or
identities), multi-object state estimation provides a state-space formulation
of the \emph{multi-object tracking} problem where the aim is to estimate the
number of objects and their trajectories \cite{Blackman, Bar88, Mahler07}.
Numerically, this problem is far more complex than standard state estimation
due to additional challenges such as false measurements, misdetection and data
association uncertainty.

In multi-object state estimation, the\ labeled multi-object filtering
recursion admits an analytic solution known as the Generalized Labeled
Multi-Bernoulli (GLMB) filter \cite{VoGLMB13}, \cite{VVP_GLMB13}. Moreover,
this recursion can be implemented with linear complexity in the number of
measurements and quadratic in the number of hypothesized objects \cite{VVH17}.
Since the filtering density only considers information on the current
multi-object state, earlier estimates cannot be updated with current data.
Consequently, apart from poorer performance compared to smoothing, an
important drawback in a multi-object context is \emph{track fragmentation},
where terminated trajectories are picked up again as new evidence from the
data emerges.
%discussion on forwardbackward smoothing;... only considers the marginals at a given time,
%hence estimate is only best for a given instance rather than on a window.
%Multi-object smoothing has been considered in the PHD, CPHD forward-backward smoother [Kirubar, Mahler, Vo, Clark] and the forward-backward GLMB smoother [Beard].
%However, the forward-backward smoothing strategy only considers the multi-object state at a given time,
%and does not necessarily eliminate track fragmentation.

In this paper, we extend the GLMB filtering recursion to a (labeled)
multi-object posterior recursion. Such posterior captures all information on
the set of underlying trajectories and eliminates track fragmentation as well
as improving general tracking performance. Specifically, by introducing the
multi-scan GLMB model, an analytic multi-object posterior recursion is
derived. Interestingly, the multi-scan GLMB recursion takes on an even simpler
and more intuitive form than the GLMB recursion. In implementation, however,
the multi-scan GLMB recursion is far more challenging. Like the GLMB filter,
the multi-scan GLMB filter needs to be truncated, and as shown in this
article, truncation by retaining components with highest weights minimizes the
$L_{1}$ truncation error. Unlike the GLMB filter, finding the significant
components of a multi-scan GLMB filter is an NP-hard multi-dimensional
assignment problem. To solve this problem, we propose an extension of the
Gibbs sampler for the 2-D assignment problem in \cite{VVH17} to higher
dimensions. The resulting technique can be applied to compute the GLMB
posterior off-line in one batch, or recursively as new observations arrive,
thereby performing smoothing-while-filtering.

The remainder of this article is divided into 5 Sections. Section
\ref{sec:background} summarizes relevant concepts in Bayesian multi-object
state estimation and the GLMB filter. Section \ref{sec:PosteriorGLMB}
introduces the multi-scan GLMB model and the multi-scan GLMB posterior
recursion. Section \ref{sec:Implementation} presents an implementation of the
multi-scan GLMB recursion using Gibbs sampling. Numerical studies are
presented in Section \ref{sec:Numericals} and conclusions are given in Section
\ref{sec:Conclusions}.

\section{Background$\label{sec:background}$}

Following the convention in \cite{VoGLMB13}, the list of variables
$X_{m},X_{m+1},...,X_{n}$ is abbreviated as $X_{m:n}$, and the inner product
$\int f(x)g(x)dx$ is denoted by $\left\langle f,g\right\rangle $. For a given
set $S$, $1_{S}(\cdot)$ denotes the indicator function of $S$, and
$\mathcal{F}(S)$ denotes the class of finite subsets of $S$. For a finite set
$X$, its cardinality (or number of elements) is denoted by $|X|$, and the
product $%
%TCIMACRO{\tprod _{x\in X}}%
%BeginExpansion
{\textstyle\prod_{x\in X}}
%EndExpansion
f(x)$, for some function $f$, is denoted by the multi-object exponential
$f^{X}$, with $f^{\emptyset}=1$. In addition we use
\[
\delta_{Y}[X]\triangleq\left\{
\begin{array}
[c]{l}%
1,\text{ if }X=Y\\
0,\text{ otherwise}%
\end{array}
\right.
\]
for a generalization of the Kroneker delta that takes arbitrary arguments.

\subsection{Trajectories and Multi-object States$\label{subsec:trajectories}$}

This subsection summarizes the representation of trajectories via labeled
multi-object states.

At time $k$, an existing object is described by a vector $x\in\mathbb{X}$ and
a unique label $\ell=(s,\alpha)$, where $s$ is the \textit{time of birth}, and
$\alpha$ is a unique index to distinguish objects born at the same time (see
Fig. 1 in \cite{VVP_GLMB13}). Let $\mathbb{B}_{s}$ denote the label space for
objects born at time $s$, then the label space for all objects up to time $k$
(including those born prior to $k$) is given by the disjoint union
$\mathbb{L}_{k}=%
%TCIMACRO{\tbiguplus \nolimits_{s=0}^{k}}%
%BeginExpansion
{\textstyle\biguplus\nolimits_{s=0}^{k}}
%EndExpansion
\mathbb{B}_{s}$ (note that $\mathbb{L}_{k}=\mathbb{L}_{k-1}\uplus
\mathbb{B}_{k}$). Hence, a \emph{labeled state} $\mathbf{x=}(x,\ell)$ at time
$k$ is an element of $\mathbb{X}\mathcal{\times}\mathbb{L}_{k}$.

A \emph{trajectory} is a sequence of labeled states with a common label, at
consecutive times \cite{VoGLMB13}, i.e. a trajectory with label $\ell
=(s,\alpha)$ and kinematic states $x_{s},x_{s+1},...,x_{t}\in\mathbb{X},$ is
the sequence
\begin{equation}
\mathbf{\tau}=[(x_{s},\ell),(x_{s+1},\ell),...,(x_{t},\ell
)].\label{eq:Trajectory0}%
\end{equation}

A \emph{labeled multi-object} state at time $i$ is a finite subset
$\mathbf{X}$ of $\mathbb{X}\mathcal{\times}\mathbb{L}_{i}$ with \emph{distinct
labels}. More concisely, let $\mathcal{L}:\mathbb{X}\mathcal{\times}%
\mathbb{L}_{k}\rightarrow\mathbb{L}_{k}$ be the projection defined by
$\mathcal{L}((x,\ell))=\ell$, then $\mathbf{X}$ has distinct labels if and
only if the \emph{distinct label indicator} $\Delta(\mathbf{X})\triangleq
\delta_{|\mathbf{X}|}[|\mathcal{L(}\mathbf{X})|]$ equals one. The labeled
states, at time $i$, of a set $S$ of trajectories (with distinct labels) is
the labeled multi-object state $\mathbf{X}_{i}=\left\{  \mathbf{\tau
(}i\mathbf{)}:\mathbf{\tau\in}S\right\}  $, where $\mathbf{\tau(}i\mathbf{)}$
denotes the labeled state of trajectory $\mathbf{\tau}$ at time $i$.

Consider a sequence $\mathbf{X}_{j:k}$ of labeled multi-object states in the
interval $\{j:k\}$. Let $\mathbf{x}_{i}^{(\mathbf{\ell})}=(x_{i}%
^{(\mathbf{\ell})},\mathbf{\ell})$ denote the element of $\mathbf{X}_{i}$ with
label $\ell\in\mathcal{L}(\mathbf{X}_{i}\mathbf{)}$. Then the trajectory in
$\mathbf{X}_{j:k}$ with label $\ell\in\cup_{i=j}^{k}\mathcal{L}(\mathbf{X}%
_{i}\mathbf{)}$ is the sequence of states with label $\ell$:
\begin{equation}
\mathbf{x}_{s(\mathbf{\ell}):t(\mathbf{\ell})}^{(\mathbf{\ell})}%
=[(x_{s(\mathbf{\ell})}^{(\mathbf{\ell})},\mathbf{\ell}%
),...,(x_{t(\mathbf{\ell})}^{(\mathbf{\ell})},\mathbf{\ell}%
)],\label{eq;Trajectory}%
\end{equation}
where%
\begin{equation}
s(\mathbf{\ell})=\max\{j,\ell\lbrack1,0]^{T}\}
\end{equation}
is the start time of label $\mathbf{\ell}$ in the interval $\{j:k\}$, and%
\begin{equation}
t(\mathbf{\ell})=s(\mathbf{\ell})+\sum_{i=s(\mathbf{\ell})+1}^{k}%
1_{\mathcal{L}(\mathbf{X}_{i}\mathbf{)}}(\mathbf{\ell})
\end{equation}
is the latest time in $\{s(\mathbf{\ell}):k\}$ such that label $\mathbf{\ell}$
still exists.

The multi-object state sequence $\mathbf{X}_{j:k}$ can thus be equivalently
represented by the set of all such trajectories, i.e.%
\begin{equation}
\mathbf{X}_{j:k}\equiv\left\{  \mathbf{x}_{s(\mathbf{\ell}):t(\mathbf{\ell}%
)}^{(\mathbf{\ell})}:\ell\in%
%TCIMACRO{\tbigcup _{i=j}^{k}}%
%BeginExpansion
{\textstyle\bigcup_{i=j}^{k}}
%EndExpansion
\mathcal{L}(\mathbf{X}_{i}\mathbf{)}\right\}  .\label{eq:SOT}%
\end{equation}
The left and right hand sides of (\ref{eq:SOT}) are simply different groupings
of the labeled states on the interval $\{j:k\}$. The multi-object state
sequence groups the labeled states according to time while the set of
trajectories groups according to labels (see also figure 1 of
\cite{VVP_GLMB13}).

For the rest of the article, single-object states are represented by lowercase
letters (e.g. $x$, $\mathbf{x}$), while multi-object states are represented by
uppercase letters (e.g. $X$, $\mathbf{X}$), symbols for labeled states and
their distributions are bolded to distinguish them from unlabeled ones (e.g.
$\mathbf{x}$, $\mathbf{X}$, $\mathbf{\pi}$, etc).

\subsection{Bayes recursion$\label{subsec:Bayes}$}

Following the Bayesian paradigm, each labeled multi-object state is modeled as
a labeled random finite set (RFS) \cite{VoGLMB13}, characterized by the Finite
Set Statistic (FISST) multi-object density \cite{MahlerPHD}, \cite{VSD05}.

Given the observation history $Z_{1:k}=(Z_{1},...,Z_{k})$, all information on
the set of objects (and their trajectories) is captured in the
\emph{multi-object posterior density, }$\mathbf{\pi}_{0:k}(\mathbf{X}%
_{0:k})\triangleq\mathbf{\pi}_{0:k}(\mathbf{X}_{0:k}|Z_{1:k})$. Note that the
dependence on $Z_{1:k}$ is omitted for notational compactness. Similar to
standard Bayesian state estimation \cite{Briers08},\cite{Doucetoverview}, the
(multi-object) posterior density can be propagated forward recursively by%
\begin{equation}
\mathbf{\pi}_{0:k}(\mathbf{X}_{0:k})=\frac{g_{k}(Z_{k}|\mathbf{X}%
_{k})\mathbf{f}_{k|k-1}(\mathbf{X}_{k}|\mathbf{X}_{k-1})\mathbf{\pi}%
_{0:k-1}(\mathbf{X}_{0:k-1})}{h_{k}(Z_{k}|Z_{1:k-1})},\label{eq:MO_posterior}%
\end{equation}
where $g_{k}(\cdot|\cdot)$ is the \emph{multi-object likelihood}
\emph{function} at time $k$, $\mathbf{f}_{k|k-1}(\cdot|\cdot)$ is the
\emph{multi-object transition density} to time $k$, and $h_{k}(Z_{k}%
|Z_{1:k-1})$ is the normalising constant, also known as the predictive
likelihood. A valid $\mathbf{f}_{k|k-1}(\cdot|\cdot)$ ensures each surviving
object keeps the same label and dead labels never reappear \cite{VoGLMB13}, so
that the multi-object history $\mathbf{X}_{0:k}$ represents a set of trajectories.

Markov Chain Monte Carlo (MCMC) approximations of the posterior have been
proposed in \cite{Vu14} and \cite{Craciun} for detection and image
measurements respectively. Combining MCMC with the generic multi-object
particle filter \cite{VSD05} has also been suggested in \cite{Kim16}.

A cheaper alternative is the \emph{multi-object filtering density},
$\mathbf{\pi}_{k}(\mathbf{X}_{k})\triangleq\int\mathbf{\pi}_{0:k}%
(\mathbf{X}_{0:k})\delta\mathbf{X}_{0:k-1}$, which can be propagated by the
\emph{multi-object Bayes filter} \cite{MahlerPHD, Mahler07}
\begin{equation}
\mathbf{\pi}_{k}(\mathbf{X}_{k})=\frac{g_{k}(Z_{k}|\mathbf{X}_{k}%
)\int\mathbf{f}_{k|k-1}(\mathbf{X}_{k}|\mathbf{X}_{k-1})\mathbf{\pi}%
_{k-1}(\mathbf{X}_{k-1})\delta\mathbf{X}_{k-1}}{h_{k}(Z_{k}|Z_{1:k-1})}.
\label{eq:MO_filtering}%
\end{equation}
Under the standard multi-object system model, the filtering recursion
(\ref{eq:MO_filtering}) admits an analytic solution known as the
\emph{Generalized labeled Multi-Bernoulli} (GLMB) filter \cite{VoGLMB13},
\cite{VVP_GLMB13}. For a general system model, the generic multi-object
particle filter can be applied, see for example \cite{PapiKim15}.

\subsection{Multi-object System model$\label{subsec:systemmodel}$}

Given a multi-object state $\mathbf{X}_{k-1}$ (at time $k-1$), each state
$\mathbf{x}_{k-1}=(x_{k-1},\ell_{k-1})\in\mathbf{X}_{k-1}$ either survives
with probability $P_{S,k-1}(\mathbf{x}_{k-1})$ and evolves to a new state
$(x_{k},\ell_{k})$ with probability density $f_{S,k|k-1}(x_{k}|x_{k-1}%
,\ell_{k-1})\delta_{\ell_{k-1}}[\ell_{k}]$ or dies with probability
$Q_{S,k-1}(\mathbf{x}_{k-1})=1-P_{S,k-1}(\mathbf{x}_{k-1})$. Further, for each
$\ell_{k}$ in a (finite) birth label space $\mathbb{B}_{k}$ at time $k$,
either a new object with state $(x_{k},\ell_{k})$ is born with probability
$P_{B,k}(\ell_{k})$ and density $f_{B,k}(x_{k},\ell_{k})$, or unborn with
probability $Q_{B,k}(\ell_{k})=1-P_{B,k}(\ell_{k})$. The multi-object state
$\mathbf{X}_{k}$\ (at time $k$) is the superposition of surviving states and
new born states, and the multi-object transition density $\mathbf{f}%
_{k|k-1}(\mathbf{X}_{k}|\mathbf{X}_{k-1})$ is given by equation (6) in
\cite{VVP_GLMB13}. An alternative form (using multi-scan exponential notation
introduced in the next section) is given in subsection \ref{sec:MultiScanGLMB}.

Given a multi-object state $\mathbf{X}_{k}$, each $\mathbf{x}_{k}\in
\mathbf{X}_{k}$ is either detected with probability $P_{D,k}(\mathbf{x}_{k})$
and generates a detection $z$ with likelihood $g_{D,k}(z|\mathbf{x}_{k})$ or
missed with probability $Q_{D,k}(\mathbf{x}_{k})=1-P_{D,k}(\mathbf{x}_{k})$.
The \emph{multi-object observation }$Z_{k}$ is the superposition of the
observations from detected objects and Poisson clutter with intensity
$\kappa_{k}$. Assuming that, conditional on $\mathbf{X}_{k}$, detections are
independent of each other and clutter, the multi-object likelihood function is
given by \cite{VoGLMB13}, \cite{VVP_GLMB13}%
\[
g_{k}(Z_{k}|\mathbf{\mathbf{X}}_{k})\propto\sum_{\theta_{k}\in\Theta_{k}%
}1_{\Theta_{k}(\mathcal{L(}\mathbf{X}_{k}))}(\theta_{k})\left[  \psi
_{k,\!Z_{k}}^{(\theta_{k}\circ\mathcal{L}(\cdot))}(\cdot)\right]
^{\mathbf{X}_{k}}%
\]
where
\[
\psi_{k,\{z_{1},...,z_{m}\}}^{(j)}(x,\ell)=\left\{
\begin{array}
[c]{ll}%
\frac{P_{D,k}(x,\ell)g_{D,k}(z_{j}|x,\ell)}{\kappa_{k}(z_{j})}, & \text{if
}j>0\\
Q_{D,k}(x,\ell) & \text{if }j=0
\end{array}
\right.  ,
\]
$\Theta_{k}$ denotes the set of \emph{positive 1-1} maps (i.e. those that
never \emph{assign distinct arguments to the same positive value}) from
$\mathbb{L}_{k}$ to $\{0$:$|Z_{k}|\}$, and $\Theta_{k}(I)$ denotes the subset
of $\Theta_{k}$ with domain $I$. The map $\theta_{k}$ assigns a detected label
$\ell$ to measurement $z_{\theta_{k}(\ell)}\in Z_{k}$, while for an undetected
label $\theta_{k}(\ell)=0$.

\subsection{GLMB Filtering recursion$\label{subsec:GLMBfilter}$}

Given a state space $\mathbb{X}$ and a discrete space $\mathbb{L}$, a
\emph{generalized labeled multi-Bernoulli} (GLMB) density on $\mathcal{F}%
(\mathbb{X\times L})$ has the form \cite{VoGLMB13}:%
\begin{equation}
\mathbf{\pi}\mathcal{(}\mathbf{X})=\Delta(\mathbf{X})\sum_{\xi\in\Xi}w^{(\xi
)}(\mathcal{L(}\mathbf{X}))\left[  p^{(\xi)}\right]  ^{\mathbf{X}%
},\label{eq:GLMB}%
\end{equation}
where $\Xi$ is a discrete index set, each $p^{(\xi)}(\cdot,\ell)$ is a
probability density on $\mathbb{X}$, i.e., $\int p^{(\xi)}(x,\ell)dx=1$, and
each $w^{(\xi)}(L)$ is non-negative with $\sum_{L\subseteq\mathbb{L}}\sum
_{\xi\in\Xi}w^{(\xi)}(L)=1$. The GLMB density (\ref{eq:GLMB}) can be
interpreted as a mixture of (labeled) multi-object exponentials.

The GLMB family is closed under the Bayes multi-object filtering recursion
(\ref{eq:MO_filtering}) and an explicit expression relating the filtering
density at time $k$ to that at time $k-1$ is given by (14) of \cite{VVH17}.
This recursion can be expressed in the following form\footnote{This involves a
straight forward change of notation, but for completeness, details are given
in Appendix \ref{sec:AppendixAltGLMB}.}, which complies with, and facilitates
the generalisation to the posterior recursion:

Given the GLMB filtering density
\begin{equation}
\mathbf{\pi}_{k-1}(\mathbf{X}_{k-1})=\Delta(\mathbf{X}_{k-1})\sum_{\xi}%
w_{k-1}^{(\xi)}(\mathcal{L(}\mathbf{X}_{k-1}))\left[  p_{k-1}^{(\xi)}\right]
^{\mathbf{X}_{k-1}}, \label{eq:delta-glmb}%
\end{equation}
\bigskip\ at time $k-1$, GLMB filtering density at time $k$ is given by%
\begin{equation}
\mathbf{\pi}_{k}(\mathbf{X}_{k})\!\propto\Delta\!(\mathbf{X}_{k})\!\sum
_{\xi,\theta_{k},I_{k-1}}\!w_{k}^{(\xi,\theta_{k})}(I_{k-1})\delta
_{\mathcal{D(}\theta_{k})}[\mathcal{L}(_{\!}\mathbf{X}_{k\!})]\!\left[
p_{k}^{(\xi,\theta_{k})}\right]  ^{\mathbf{X}_{k}}\! \label{eq:GLMB_joint0}%
\end{equation}
where $\xi\in\Xi$, $\theta_{k}\in\Theta_{k}$, $I_{k-1}\in\mathcal{F}%
(\mathbb{L}_{k-1})$, $\mathcal{D}(\theta_{k})$ denotes the domain of
$\theta_{k}$, \allowdisplaybreaks%
\begin{align}
w_{k}^{(\xi,\theta_{k})}(I_{k-1})  &  =1_{\mathcal{F}(\mathbb{B}%
_{k}\mathbf{\uplus}I_{k-1})}(\mathcal{D(}\theta_{k}))\left[  w_{k|k-1}%
^{(\xi,\theta_{k})}\right]  ^{\mathbb{B}_{k}\uplus I_{k-1}}w_{k-1}^{(\xi
)}(I_{k-1})\label{eq:GLMB_joint1}\\
w_{k|k-1}^{(\xi,\theta_{k})}(\ell)  &  =\left\{  \!%
\begin{array}
[c]{ll}%
\!\bar{\Lambda}_{B,k}^{(\theta_{k}(\ell))}(\ell), & \!\ell\in\mathcal{D(}%
\theta_{k})\cap\mathbb{B}_{k},\\
\!\bar{\Lambda}_{S,k|k-1}^{(\xi,\theta_{k}(\ell))}(\ell), & \!\ell
\in\mathcal{D(}\theta_{k})-\mathbb{B}_{k},\\
\!Q_{B,k}(\ell), & \!\ell\in\mathbb{B}_{k}-\mathcal{D(}\theta_{k}),\\
\!\bar{Q}_{S,k-1}^{(\xi)}(\ell), & \!\text{otherwise},
\end{array}
\right.  ,\\
p_{k}^{(\xi,\theta_{k})}(x,\ell)  &  =\left\{  \!%
\begin{array}
[c]{ll}%
\!\frac{\Lambda_{\!B,k}^{(\theta_{k}(\ell))}(x,\ell)}{\bar{\Lambda}%
_{B,k}^{(\theta_{k}(\ell))}(\ell)}, & \!\ell\in\mathcal{D(}\theta_{k}%
)\cap\mathbb{B}_{k},\\
\!\frac{\left\langle \Lambda_{\!S,k|k-1}^{(\theta_{k}(\ell))}(x|\cdot
,\ell),p_{k-1}^{(\xi)}(\cdot,\ell)\right\rangle }{\bar{\Lambda}_{S,k|k-1}%
^{(\xi,\theta_{k}(\ell))}(\ell)}, & \!\ell\in\mathcal{D(}\theta_{k}%
)-\mathbb{B}_{k},
\end{array}
\right.  ,\label{eq:GLMB_joint2}\\
\Lambda_{\!B,k}^{(j)}(x,\ell)  &  =\psi_{\!k,Z_{k}}^{(j)}(x,\ell
)f_{B,k}(x,\ell)P_{B,k}(\ell),\label{eq:GLMB_joint3}\\
\Lambda_{\!S,k|k-1}^{(j)}(x|\varsigma,\ell)  &  =\psi_{\!k,Z_{k}}^{(j)}%
(x,\ell)f_{S.k|k-1}(x|\varsigma,\ell)P_{S,k-1}(\varsigma,\ell
),\label{eq:GLMB_joint4}\\
\bar{Q}_{S,k-1}^{(\xi)}(\ell)  &  =\left\langle Q_{S,k-1}(\cdot,\ell
),p_{k-1}^{(\xi)}(\cdot,\ell)\right\rangle ,\label{eq:GLMB_joint5}\\
\bar{\Lambda}_{B,k}^{(j)}(\ell)  &  =\left\langle \Lambda_{\!B,k}^{(j)}%
(\cdot,\ell),1\right\rangle ,\label{eq:GLMB_joint6}\\
\bar{\Lambda}_{S.k|k-1}^{(\xi,j)}(\ell)  &  =\int\left\langle \Lambda
_{\!S,k|k-1}^{(j)}(x|\cdot,\ell),p_{k-1}^{(\xi)}(\cdot,\ell)\right\rangle dx.
\label{eq:GLMB_joint7}%
\end{align}

\vspace{8pt}

The number of components of the GLMB filtering density grows
super-exponentially in time. Truncation by discarding components with small
weights minimizes the $L_{1}$ approximation error in the multi-object density
\cite{VVP_GLMB13}. This can be achieved by solving the ranked assignment
problems using Murty's algorithm or Gibbs sampling \cite{VVH17}.

\section{GLMB Posterior Recursion$\label{sec:PosteriorGLMB}$}

In this section we extend the GLMB model to the multi-scan case, and
subsequently derive an analytic recursion for the multi-scan GLMB posterior.

\subsection{Multi-scan GLMB$\label{sec:MultiScanGLMB}$}

Recall the equivalence between the multi-object state sequence $\mathbf{X}%
_{j:k}$ and the set $\left\{  \mathbf{x}_{s(\mathbf{\ell}):t(\mathbf{\ell}%
)}^{(\mathbf{\ell})}:\ell\in\cup_{i=j}^{k}\mathcal{L}(\mathbf{X}_{i}%
\mathbf{)}\right\}  $ of trajectories in (\ref{eq:SOT}). For any function $h$
taking the trajectories to the non-negative reals we introduce the following
so-called \emph{multi-scan exponential} notation:%
\begin{equation}
\left[  h\right]  ^{\mathbf{X}_{j:k}}\triangleq\left[  h\right]  ^{\left\{
\mathbf{x}_{s(\mathbf{\ell}):t(\mathbf{\ell})}^{(\mathbf{\ell})}:\ell\in
\cup_{i=j}^{k}\mathcal{L}(\mathbf{X}_{i}\mathbf{)}\right\}  }=\prod
\limits_{\ell\in\cup_{i=j}^{k}\mathcal{L}(\mathbf{X}_{i}\mathbf{)}%
}h(\mathbf{x}_{s(\mathbf{\ell}):t(\mathbf{\ell})}^{(\mathbf{\ell}%
)})\label{eq:MSMOexp}%
\end{equation}
Note from (\ref{eq;Trajectory}) that a trajectory $\mathbf{x}_{s(\mathbf{\ell
}):t(\mathbf{\ell})}^{(\mathbf{\ell})}$ is completely characterised by $\ell$
and the kinematic states $x_{s(\mathbf{\ell}):t(\mathbf{\ell})}^{(\ell)}$,
hence we write $h(\mathbf{x}_{s(\mathbf{\ell}):t(\mathbf{\ell})}%
^{(\mathbf{\ell})})$ and $h(x_{s(\mathbf{\ell}):t(\mathbf{\ell})}^{(\ell
)};\ell)$ interchangeably.

The multi-scan exponential notation is quite suggestive since $\left[
gh\right]  ^{\mathbf{X}_{j:k}}=\left[  g\right]  ^{\mathbf{X}_{j:k}}\left[
h\right]  ^{\mathbf{X}_{j:k}}$, and if the labels of $\mathbf{X}_{j:k}$,
$\mathbf{Y}_{j:k}$ are disjoint then $\left[  h\right]  ^{\mathbf{X}%
_{j:k}\uplus\mathbf{Y}_{j:k}}=\left[  h\right]  ^{\mathbf{X}_{j:k}}\left[
h\right]  ^{\mathbf{Y}_{j:k}}$ (see Appendix \ref{sec:AppendixProperties} for
additional properties). It also provides an intuitive expression for the
multi-object transition density in \cite{VoGLMB13}.

\begin{proposition}
$\label{prop:transition}$ For the multi-object dynamic model described in
subsection \ref{subsec:systemmodel}, the multi-object transition density is
given by%
\begin{equation}
\mathbf{f}_{k|k-1}\left(  \mathbf{\mathbf{X}}_{k}|\mathbf{X}_{k-1}\right)
=\Delta(\mathbf{\mathbf{X}}_{k})1_{\mathcal{F}(\mathbb{B}_{k}\mathbf{\uplus
}\mathcal{L(}\mathbf{X}_{k-1}))}(\mathcal{L(}\mathbf{X}_{k}))Q_{B,k}%
^{\mathbb{B}_{k}-\mathcal{L(}\mathbf{\mathbf{X}}_{k})}\left[  \phi
_{k-1:k}\right]  ^{\mathbf{X}_{k-1:k}}\label{eq:labeled_transition_MS}%
\end{equation}
where%
\begin{equation}
\phi_{k-1:k}(x_{s(\ell):t(\ell)}^{(\ell)};\ell)=\left\{
\begin{array}
[c]{ll}%
P_{B,k}(\ell)f_{B,k}(x_{k}^{(\ell)},\ell) & s(\ell)=k\\
P_{S,k-1}(x_{k-1}^{(\ell)},\ell)f_{S,k|k-1}(x_{k}^{(\ell)}|x_{k-1}^{(\ell
)},\ell) & t(\ell)=k>s(\ell)\\
Q_{S,k-1}(x_{k-1}^{(\ell)},\ell), & t(\ell)=k-1
\end{array}
\right.  \label{eq:labeled_transition_MS1}%
\end{equation}
For completeness the proof is given in Appendix \ref{sec:AppendixTransition}.
\end{proposition}

\begin{definition}
A \emph{multi-scan GLMB} density on $\mathcal{F}(\mathbb{X\times L}_{j}%
)\times...\times\mathcal{F}(\mathbb{X\times L}_{k})$ is defined by
\begin{equation}
\mathbf{\pi}(\mathbf{X}_{j:k})=\Delta(\mathbf{X}_{j:k})\sum_{\xi\in\Xi}%
w^{(\xi)}(\mathcal{L(}\mathbf{X}_{j:k}))\left[  p^{(\xi)}\right]
^{\mathbf{X}_{j:k}} \label{eq:MSGLMB}%
\end{equation}
where: $\Xi$ is a discrete index set; $\Delta(\mathbf{X}_{j:k})\triangleq%
%TCIMACRO{\tprod \nolimits_{i=j}^{k}}%
%BeginExpansion
{\textstyle\prod\nolimits_{i=j}^{k}}
%EndExpansion
\Delta(\mathbf{X}_{i})$; $\mathcal{L(}\mathbf{X}_{j:k})\triangleq
\lbrack\mathcal{L(}\mathbf{X}_{j}),...,\mathcal{L(}\mathbf{X}_{k})]$; each
$w^{(\xi)}(I_{j:k})$, $\xi\in\Xi$, $I_{j:k}\in\mathcal{F}(\mathbb{L}%
_{j})\times...\times\mathcal{F}(\mathbb{L}_{k})$ is non-negative with%
\begin{equation}
\sum_{\xi}\sum_{I_{j:k}}w^{(\xi)}(I_{j:k})=1; \label{eq:MSGLMBweight}%
\end{equation}
and each $p^{(\xi)}(\cdot;\ell)$, $\xi\in\Xi$, $\ell\in\cup_{i=j}%
^{k}\mathcal{L}(\mathbf{X}_{i}\mathbf{)}$ is a probability density on
$\mathbb{X}^{t(\ell)-s(\ell)+1}$, i.e.,%
\begin{equation}
\int p^{(\xi)}(x_{s(\mathbf{\ell}):t(\mathbf{\ell})};\ell)dx_{s(\mathbf{\ell
}):t(\mathbf{\ell})}=1. \label{eq:MSGLMBtrackdensity}%
\end{equation}

\end{definition}

It is clear that the multi-scan GLMB (density) reduces to a GLMB (density)
when $j=k$.

Similar to the GLMB, the multi-scan GLMB (\ref{eq:MSGLMB}) can be expressed in
the so-called $\delta$-form:%
\begin{equation}
\mathbf{\pi}(\mathbf{X}_{j:k})=\Delta(\mathbf{\mathbf{X}}_{j:k})\sum_{\xi}%
\sum_{I_{j:k}}w^{(\xi)}(I_{j:k})\delta_{I_{j:k}}[\mathcal{L(}%
\mathbf{\mathbf{X}}_{j:k})]\left[  p^{(\xi)}\right]  ^{\mathbf{\mathbf{X}%
}_{j:k}} \label{eq:d-MSGLMB}%
\end{equation}
where $\xi\in$ $\Xi$, $I_{j:k}\in\mathbf{\mathbf{\mathcal{F}(}\mathbb{L}}%
_{j}\mathbf{\mathbb{)}\mathcal{\times}}...\mathbf{\mathcal{\times F}%
(}\mathbb{L}_{k}\mathbb{)}$. Each term or component of a multi-scan GLMB
consists of a weight $w^{(\xi)}(I_{j:k})$ and a multi-scan exponential
$\left[  p^{(\xi)}\right]  ^{\mathbf{\mathbf{X}}_{j:k}}$ with label history
that matches $I_{j:k}$. The\ weight $w^{(\xi)}(I_{j:k})$ can be interpreted as
the probability of hypothesis $(\xi,I_{j:k})$, and for each $\ell\in\cup
_{i=j}^{k}\mathcal{L}(I_{i}\mathbf{)}$, $p^{(\xi)}(x_{s(\mathbf{\ell
}):t(\mathbf{\ell})};\ell)$ is the joint probability density of its kinematic
states, given hypothesis $(\xi,I_{j:k})$.

\begin{proposition}
$\label{prop:expectation}$The integral of a function $f:$ $\mathcal{F}%
(\mathbb{L}_{j})\times...\times\mathcal{F}(\mathbb{L}_{k})\rightarrow
\mathbb{R}$ with respect to the multi-scan GLMB (\ref{eq:MSGLMB}) is
\begin{equation}
\int f(\mathcal{L(}\mathbf{\mathbf{X}}_{j:k}))\mathbf{\pi}(\mathbf{X}%
_{j:k})\delta\mathbf{X}_{j:k}=\sum_{\xi}\sum_{I_{j:k}}f(I_{j:k})w^{(\xi
)}(I_{j:k})
\end{equation}
where $\xi\in$ $\Xi$, $I_{j:k}\in\mathbf{\mathbf{\mathcal{F}(}\mathbb{L}}%
_{j}\mathbf{\mathbb{)}\mathcal{\times}}...\mathbf{\mathcal{\times F}%
(}\mathbb{L}_{k}\mathbb{)}$. See Appendix \ref{sec:AppendixExpectation} for proof.
\end{proposition}

By setting $f$ to 1 in the above proposition, the multi-scan GLMB integrates
to 1, and hence, is a FISST density. Some useful statistics from the
multi-scan GLMB follows from the above proposition for suitably defined
functions of the labels.

\begin{corollary}
The cardinality distribution, i.e. distribution of the number of trajectories
is given by%
\begin{equation}
\Pr\left(  \left\vert \cup_{i=j}^{k}\mathcal{L}(\mathbf{X}_{i}\mathbf{))}%
\right\vert =n\right)  =\sum_{\xi}\sum_{I_{j:k}}\delta_{n}\left[  \left\vert
\cup_{i=j}^{k}I_{i}\mathbf{)}\right\vert \right]  w^{(\xi)}(I_{j:k})
\label{eq:cardinality}%
\end{equation}

\end{corollary}

\begin{corollary}
The joint probability of existence of a set of trajectories with labels $L$ is
given by
\begin{equation}
\Pr\left(  L\text{ exist}\right)  =\sum_{\xi}\sum_{I_{j:k}}1_{\mathcal{F}%
\left(  \cup_{i=j}^{k}I_{i}\right)  }(L)w^{(\xi)}(I_{j:k}).
\label{eq:jointexistence}%
\end{equation}
As a special case, the probability of existence of trajectory with label
$\ell$ is
\begin{equation}
\Pr\left(  \ell\text{ exists}\right)  =\sum_{\xi}\sum_{I_{j:k}}1_{\cup
_{i=j}^{k}I_{i}}(\ell)w^{(\xi)}(I_{j:k}). \label{eq:existence}%
\end{equation}

\end{corollary}

\begin{corollary}
The distribution of trajectory lengths is given by%
\begin{equation}
\Pr\left(  \text{a trajectory has length }m\right)  =\sum_{\xi}\sum_{I_{j:k}%
}\frac{w^{(\xi)}(I_{j:k})}{\left\vert \cup_{i=j}^{k}I_{i}\right\vert }%
\sum_{\ell\in\cup_{i=j}^{k}I_{i}}\delta_{m}\left[  t(\ell)-s(\ell)+1\right]  ,
\label{eq:lengthdistribution}%
\end{equation}
and the distribution of the length of trajectory with label $\ell$ is%
\begin{equation}
\Pr\left(  \text{length}(\ell)=m\right)  =\sum_{\xi}\sum_{I_{j:k}}\delta
_{m}\left[  t(\ell)-s(\ell)+1\right]  1_{\cup_{i=j}^{k}I_{i}}(\{\ell
\})w^{(\xi)}(I_{j:k}). \label{eq:lengthprob}%
\end{equation}

\end{corollary}

Similar to its single-scan counterpart, a number of estimators can be
constructed for a multi-scan GLMB. The simplest would be to find the
multi-scan GLMB component with the highest weight $w^{(\xi)}(I_{j:k})$ and
compute the most probable or expected trajectory estimate from $p^{(\xi
)}(\cdot;\ell)$ for each $\ell\in\cup_{i=j}^{k}\mathcal{L}(I_{i}\mathbf{)}$.
Alternatively, instead of the most significant, we can use the most
significant amongst components with the most probable cardinality $n^{\ast}$
(determined by maximizing the cardinality distribution (\ref{eq:cardinality})).

Another class of estimators, based on existence probabilities, can be
constructed as follows. Find the set of labels $L^{\ast}$ with highest joint
existence probability by maximizing (\ref{eq:jointexistence}). Then for each
$\ell\in L^{\ast}$ determine the most probable length $m^{\ast}$ by maximizing
(\ref{eq:lengthprob}) and compute the trajectory density%
\begin{equation}
p(x_{s(\mathbf{\ell}):s(\mathbf{\ell})+m^{\ast}-1};\ell)\propto\sum_{\xi}%
\sum_{I_{j:k}}\delta_{m^{\ast}}\left[  t(\ell)-s(\ell)+1\right]  1_{\cup
_{i=j}^{k}I_{i}}(\{\ell\})w^{(\xi)}(I_{j:k})p^{(\xi)}(x_{s(\mathbf{\ell
}):s(\mathbf{\ell})+m^{\ast}-1};\ell),
\end{equation}
from which the most probable or expected trajectory estimate can be
determined. Alternatively, instead of the label set with highest joint
existence probability, we can use the label set of cardinality $n^{\ast}$ with
highest joint existence probability. Another option is to find the $n^{\ast}$
labels with highest individual existence probabilities and use the same
strategy for computing the trajectory estimates.

\subsection{Multi-scan GLMB Posterior Recursion $\label{subsec:LG_GMP}$}

Just as the GLMB is closed under the filtering recursion
(\ref{eq:MO_filtering}), the multi-scan GLMB is closed under the posterior
recursion (\ref{eq:MO_posterior}). Moreover, the multi-scan GLMB posterior
recursion is, in essence, the GLMB filtering recursion without the
marginalization of past labels and kinematic states. This is stated more
concisely in the following Proposition (see Appendix
\ref{sec:AppendixMultiScanRec} proof).

\begin{proposition}
$\label{prop:MSGLMBrecursion}$Under the standard multi-object system model, if
the multi-object posterior at time $k-1$ is a multi-scan GLMB of the form
\begin{equation}
\mathbf{\pi}_{0:k-1}(\mathbf{X}_{0:k-1})=\Delta(\mathbf{\mathbf{X}}%
_{0:k-1})\sum_{\xi}w_{0:k-1}^{(\xi)}(\mathcal{L(}\mathbf{X}_{0:k-1}))\left[
p_{0:k-1}^{(\xi)}\right]  ^{\mathbf{X}_{0:k-1}}, \label{eq:MSGLMBprev}%
\end{equation}
where $\xi\in\Xi$, then the multi-object posterior at time $k$ is the
multi-scan GLMB:%
\begin{equation}
\mathbf{\pi}_{0:k}(\mathbf{X}_{0:k})\propto\Delta(\mathbf{\mathbf{X}}%
_{0:k})\sum_{\xi,\theta_{k}}w_{0:k}^{(\xi,\theta_{k})}(\mathcal{L(}%
\mathbf{X}_{0:k-1}))\delta_{\mathcal{D(}\theta_{k})}[\mathcal{L(}%
\mathbf{X}_{k})]\left[  p_{0:k}^{(\xi,\theta_{k})}\right]  ^{\mathbf{X}_{0:k}}
\label{eq:MSGLMBupdate}%
\end{equation}
where $\theta_{k}\in\Theta_{k}$,\allowdisplaybreaks%
\begin{align}
w_{0:k}^{(\xi,\theta_{k})}(I_{0:k-1})  &  =1_{\mathcal{F}(\mathbb{B}%
_{k}\mathbf{\uplus}I_{k-1})}(\mathcal{D(}\theta_{k}))\left[  w_{k|k-1}%
^{(\xi,\theta_{k})}\right]  ^{\mathbb{B}_{k}\uplus I_{k-1}}w_{0:k-1}^{(\xi
)}(I_{0:k-1})\label{eq:MSGLMBupdate1}\\
w_{k|k-1}^{(\xi,\theta_{k})}(\ell)  &  =\left\{  \!%
\begin{array}
[c]{ll}%
\!\bar{\Lambda}_{B,k}^{(\theta_{k}(\ell))}(\ell), & \!\ell\in\mathcal{D(}%
\theta_{k})\cap\mathbb{B}_{k},\\
\!\bar{\Lambda}_{S,k|k-1}^{(\xi,\theta_{k}(\ell))}(\ell), & \!\ell
\in\mathcal{D(}\theta_{k})-\mathbb{B}_{k},\\
\!Q_{B,k}(\ell), & \!\ell\in\mathbb{B}_{k}-\mathcal{D(}\theta_{k}),\\
\!\bar{Q}_{S,k-1}^{(\xi)}(\ell), & \!\text{otherwise},
\end{array}
\right.  ,\\
p_{0:k}^{(\xi,\theta_{k})}(x_{s(\ell):t(\ell)}^{(\ell)};\ell)  &  =\left\{
\begin{array}
[c]{ll}%
\frac{\Lambda_{\!B,k}^{(\theta_{k}(\ell))}(x_{k}^{(\ell)},\ell)}{\bar{\Lambda
}_{B,k}^{(\theta_{k}(\ell))}(\ell)}, & s(\ell)=k\\
\frac{\Lambda_{S,k|k-1}^{(\theta_{k}(\ell))}(x_{k}^{(\ell)}|x_{k-1}^{(\ell
)},\ell)p_{0:k-1}^{(\xi)}(x_{s(\ell):k-1}^{(\ell)};\ell)}{\bar{\Lambda
}_{S,k|k-1}^{(\xi,\theta_{k}(\ell))}(\ell)}, & t(\ell)=k>s(\ell)\\
\frac{Q_{S,k-1}(x_{k-1}^{(\ell)},\ell)p_{0:k-1}^{(\xi)}(x_{s(\ell):k-1}%
^{(\ell)};\ell)}{\bar{Q}_{S,k-1}^{(\xi)}(\ell)}, & t(\ell)=k-1\\
p_{0:k-1}^{(\xi)}(x_{s(\ell):t(\ell)}^{(\ell)};\ell), & t(\ell)<k-1
\end{array}
\right.  . \label{eq:MSGLMBupdate2}%
\end{align}

\end{proposition}

\vspace{8pt}

Note that $p_{0:k}^{(\xi,\theta_{k})}(x_{s(\ell):t(\ell)}^{(\ell)};\ell)$ is
indeed a probability density since
\begin{align*}
\int\Lambda_{S,k}^{(\theta_{k}(\ell))}(x_{k},\ell|x_{k-1})p_{s(\ell
):k-1}^{(\xi)}(x_{s(\ell):k-1};\ell)dx_{s(\ell):k}  &  =\int\int\Lambda
_{S,k}^{(\theta_{k}(\ell))}(x_{k},\ell|x_{k-1})p_{k-1}^{(\xi)}(x_{k-1}%
;\ell)dx_{k-1}dx_{k}\\
&  =\bar{\Lambda}_{S,k,}^{(\xi,\theta_{k})}(\ell)\\
\int Q_{S,k-1}(x_{k-1},\ell)p_{0:k-1}^{(\xi)}(x_{s(\ell):k-1};\ell
)dx_{s(\ell):k-1}  &  =\int Q_{S,k-1}(x_{k-1},\ell)p_{k-1}^{(\xi)}%
(x_{k-1};\ell)dx_{k-1}\\
&  =\bar{Q}_{S,k-1}^{(\xi)}(\ell)
\end{align*}

The multi-scan GLMB posterior recursion (\ref{eq:MSGLMBprev}%
)-(\ref{eq:MSGLMBupdate}) bears remarkable resemblance to the GLMB filtering
recursion (\ref{eq:delta-glmb})-(\ref{eq:GLMB_joint0}). Indeed, the weight
increments for multi-scan GLMB and GLMB components are identical. Arguably,
the multi-scan GLMB recursion is more intuitive because it does not involve
marginalization over previous label sets nor past states of the trajectories.

The multi-scan GLMB recursion initiates trajectories for new labels, update
trajectories for surviving labels, terminates trajectories for disappearing
labels, and stores trajectories that disappeared earlier. Noting that $\ell
\in\mathcal{D(}\theta_{k})\cap\mathbb{B}_{k}$ is equivalent to $s(\ell)=k$,
initiation of trajectories for new labels is identical to that of the GLMB
filter. Noting $\ell\in\mathcal{D(}\theta_{k})-\mathbb{B}_{k}$ is equivalent
to $t(\ell)=k>s(\ell)$, the update of trajectories for surviving labels is the
same as the GLMB filter, but without marginalization of past kinematic states.
On the other hand, termination/storing of trajectories for
disappearing/disappeared labels are not needed in the GLMB filter.

\subsection{Cannonical Multi-scan GLMB Posterior}

Without summing over the labels nor integrating the probability densities of
the trajectories, the canonical expression for the multi-scan GLMB posterior
takes on a rather compact form. To accomplish this, we represent each
$\theta_{k}\in\Theta_{k}$ by an extended association map $\gamma
_{k}:\mathbb{L}_{k}\rightarrow\{-1$:$|Z_{k}|\}$ defined by%
\begin{equation}
\gamma_{k}(\ell)=\left\{
\begin{array}
[c]{ll}%
\theta_{k}(\ell), & \text{if }\ell\in\mathcal{D(}\theta_{k})\\
-1, & \text{otherwise}%
\end{array}
\right.  .
\end{equation}
Let $\Gamma_{k}$ denote the set of positive 1-1 maps from $\mathbb{L}_{k}$ to
$\{-1$:$|Z_{k}|\}$, and (with a slight abuse of notation) denote the live
labels of $\gamma_{k}$, i.e. the domain $\mathcal{D(}\theta_{k})$, by%
\[
\mathcal{L(}\gamma_{k})\triangleq\{\ell\in\mathbb{L}_{k}:\gamma_{k}(\ell
)\geq0\}.
\]
Then for any $\gamma_{k}\in\Gamma_{k}$, we can recover $\theta_{k}\in
\Theta_{k}$ by $\theta_{k}(\ell)=\gamma_{k}(\ell)$ for each $\ell
\in\mathcal{L(}\gamma_{k})$. It is clear that there is a bijection between
$\Theta_{k}$ and $\Gamma_{k}$, and hence $\theta_{1:k}$ can be completely
represented by $\gamma_{1:k}$.

Starting with an empty initial posterior $\mathbf{\pi}_{0}(\mathbf{X}%
_{0})=\delta_{0}[\mathcal{L(}\mathbf{X}_{0})]$, by iteratively applying
Proposition \ref{prop:MSGLMBrecursion}, the posterior at time $k$ is given by
\begin{equation}
\mathbf{\pi}_{0:k}(\mathbf{X}_{0:k})\propto\Delta(\mathbf{\mathbf{X}}%
_{0:k})\sum_{\gamma_{1:k}}w_{0:k}^{(\gamma_{0:k})}\delta_{\mathcal{L(}%
\gamma_{0:k})}[\mathcal{L(}\mathbf{X}_{0:k})]\left[  p_{0:k}^{(\gamma_{0:k}%
)}\right]  ^{\mathbf{X}_{0:k}}\label{eq:MSGLMBCannonical0}%
\end{equation}
where $\mathcal{L(}\gamma_{0})\triangleq\emptyset$,\allowdisplaybreaks%
\begin{align}
w_{0:k}^{(\gamma_{0:k})} &  =\prod\limits_{i=1}^{k}1_{\Gamma_{i}}(\gamma
_{i})1_{\mathcal{F}(\mathbb{B}_{i}\mathbf{\uplus}\mathcal{L(}\gamma_{i-1}%
))}(\mathcal{L(}\gamma_{i}))\left[  \omega_{i|i-1}^{(\gamma_{0:i}(\cdot
))}\right]  ^{\mathbb{B}_{i}\uplus\mathcal{L(}\gamma_{i-1})}%
\label{eq:MSGLMBCannonical1}\\
\omega_{i|i-1}^{(\gamma_{0:i}(\ell))} &  =\left\{
\begin{array}
[c]{ll}%
\bar{\Lambda}_{B,i}^{(\gamma_{i}(\ell))}(\ell), & \ell\in\mathcal{L(}%
\gamma_{i})\cap\mathbb{B}_{i}\\
\bar{\Lambda}_{S,i|i-1}^{(\gamma_{0:i-1},\gamma_{i}(\ell))}(\ell), & \ell
\in\mathcal{L(}\gamma_{i})-\mathbb{B}_{i}\\
Q_{B,i}(\ell), & \ell\in\mathbb{B}_{i}-\mathcal{L(}\gamma_{i})\\
\bar{Q}_{S,i-1}^{(\gamma_{0:i-1})}(\ell), & \text{otherwise},
\end{array}
\right.  \label{eq:MSGLMBCannonical2}\\
p_{0:k}^{(\gamma_{0:k})}(x_{s(\ell):t(\ell)}^{(\ell)};\ell) &  \propto\left\{
\begin{array}
[c]{ll}%
\Lambda^{(\gamma_{0:k}(\ell))}(x_{s(\ell):k}^{(\ell)};\ell) & \ell
\in\mathcal{L(}\gamma_{k})\\
Q_{S,t(\ell)}(x_{t(\ell)}^{(\ell)},\ell)\Lambda^{(\gamma_{0:t(\ell)}(\ell
))}(x_{s(\ell):t(\ell)}^{(\ell)};\ell) & \text{otherwise}%
\end{array}
\right.  \label{eq:MSGLMBCannonical3}\\
\Lambda^{(\gamma_{0:j}(\ell))}(x_{s(\ell):j}^{(\ell)};\ell) &  =%
%TCIMACRO{\tprod \limits_{i=s(\ell)+1}^{j}}%
%BeginExpansion
{\textstyle\prod\limits_{i=s(\ell)+1}^{j}}
%EndExpansion
\Lambda_{S,i|i-1}^{(\gamma_{i}(\ell))}(x_{i}^{(\ell)}|x_{i-1}^{(\ell)}%
,\ell)\Lambda_{\!B,s(\ell)}^{(\gamma_{s(\ell)}(\ell))}(x_{s(\ell)}^{(\ell
)},\ell),\label{eq:MSGLMBCannonical4}%
\end{align}

%While the multi-scan GLMB posterior at each time is expressible in as a finite sum, the number of terms grows super-exponentially in time. For implementation it is necessary to truncate the sum by discarding terms or components with small weights. Indeed, this strategy minimizes the L1-norm of the approximation error, using the same lines of arguments as Proposition 5 of <cite>VVP_GLMB13</cite>.

Note that the multi-scan GLMB (\ref{eq:MSGLMBCannonical1}) is completely
parameterized by the components ($w_{0:k}^{(\gamma_{0:k})}$, $p_{0:k}%
^{(\gamma_{0:k})}$ ) with positive weights. While the multi-scan GLMB
posterior at each time is expressible as a finite sum, the number of terms
grows super-exponentially in time. For implementation it is necessary to have
an approximation with a managable number of terms. Several functional
approximation criteria, e.g. $L_{p}$-norm (of the difference) or information
theoretic divergences such as Kullback-Leibler \cite{Papi_etal15}, Renyi,
Cauchy-Schwarz \cite{HVVM15}, \cite{BVVA17} and so on, can be extended to the
multi-scan case by replacing the set integral with multiple set integrals.

Computing the multi-scan GLMB posterior first and then approximating it
(according to a certain criterion) is not tractable. The main challenge is:
given a prescibed number of terms, what is the best multi-scan GLMB
approximation of the posterior, without having to compute all of its terms.

In this work we consider the $L_{1}$-norm criterion, for which an optimal
multi-scan GLMB approximation with a prescribed number of terms can be
determined (see next section). Using the same lines of arguments as
Proposition 5 of \cite{VVP_GLMB13}, the $L_{1}$-error between a multi-scan
GLMB and its truncation is given by the following result.

\begin{proposition}
\label{Prop_L1_error}Let $\left\Vert \mathbf{f}\right\Vert _{1}\triangleq
\int\left\vert \mathbf{f}(\mathbf{X}_{j:k})\right\vert \delta\mathbf{X}_{j:k}$
denote the $L_{1}$-norm of $\mathbf{f:\mathbf{\mathcal{F}(}\mathbb{X}%
\mathcal{\times}\mathbb{L}}_{j}\mathbf{\mathbb{)}\mathcal{\times}%
}...\mathbf{\mathcal{\times F}(}\mathbb{X}\mathcal{\times}\mathbb{L}%
_{k}\mathbb{)}\rightarrow\mathbb{R}$, and for a given $\mathbb{H\subseteq}%
\Xi\mathbf{\mathcal{\times}\mathbf{\mathcal{F}(}\mathbb{L}}_{j}%
\mathbf{\mathbb{)}\mathcal{\times}}...\mathbf{\mathcal{\times F}(}%
\mathbb{L}_{k}\mathbb{)}$ let
\[
\mathbf{f}_{\mathbb{H}}\mathbf{(\mathbf{X}}_{j:k}\mathbf{)}\triangleq
\Delta(\mathbf{\mathbf{X}}_{j:k})\sum\limits_{(\xi,I_{j:k})\in\mathbb{H}%
}w^{(\xi)}(I_{j:k})\delta_{I_{j:k}}[\mathcal{L(}\mathbf{\mathbf{X}}%
_{j:k})]\left[  p^{(\xi)}\right]  ^{\mathbf{\mathbf{X}}_{j:k}}%
\]
be a multi-scan GLMB density with unnormalized weights. If $\mathbb{T\subseteq
H}$ then%
\begin{align*}
||\mathbf{f}_{\mathbb{H}}-\mathbf{f}_{\mathbb{T}}||_{1}  &  \mathbf{=}%
\sum\limits_{(\xi,I_{j:k})\in\mathbb{H-T}}w^{(\xi)}(I_{j:k}),\\
\left\Vert \frac{\mathbf{f}_{\mathbb{H}}}{||\mathbf{f}_{\mathbb{H}}||_{1}%
}-\frac{\mathbf{f}_{\mathbb{T}}}{||\mathbf{f}_{\mathbb{T}}\mathbf{||}_{1}%
}\right\Vert _{1}  &  \leq2\frac{||\mathbf{f}_{\mathbb{H}}||_{1}%
-||\mathbf{f}_{\mathbb{T}}\mathbf{||}_{1}}{||\mathbf{f}_{\mathbb{H}}||_{1}}.
\end{align*}

\end{proposition}

\vspace{8pt}

Hence, given a multi-scan GLMB, the minimum $L_{1}$-norm approximation for a
prescribed number of terms can be obtained by keeping only those with highest
weights. Furthermore, this can be accomplished, without exhaustively computing
all terms of the posterior, by solving the multi-dimensional or multi-frame
assignment problem \cite{Pierskalla68}. However, this problem is NP-hard for
more than two scans. The most popular approximate solutions are based on
Lagrangian relaxation \cite{Poor91}, \cite{Poore93}, \cite{Poore97}, which is
still computationally intensive. In this work we propose a more efficient
algorithm by extending the Gibbs sampler of \cite{VVH17} to the
multi-dimensional case.

\section{Implementation$\label{sec:Implementation}$}

In this section we consider problem of finding significant components
(characterized by their association history) of the multi-scan GLMB
(\ref{eq:MSGLMBCannonical0}) by sampling from some discrete probability
distribution $\pi$. To ensure that mostly high-weight components are selected,
$\pi$ should be constructed so that association histories with high weights
are more likely to be chosen than those with low weights. A natural choice to
set%
\begin{equation}
\pi^{(j)}(\gamma_{j}|\gamma_{0:j-1})\propto1_{\Gamma_{j}}(\gamma
_{j})1_{\mathbb{B}_{j}\uplus\mathcal{L(}\gamma_{j-1})}(\mathcal{L}(\gamma
_{j}))\left[  \omega_{j|j-1}^{(\gamma_{0:j}(\cdot))}\right]  ^{\mathbb{B}%
_{j}\uplus\mathcal{L(}\gamma_{j-1})}\label{eq:conditionals0}%
\end{equation}
with $\mathcal{L}(\gamma_{j})=\emptyset$, so that%
\begin{equation}
\pi(\gamma_{1:k})=\prod\limits_{j=1}^{k}\pi^{(j)}(\gamma_{j}|\gamma
_{0:j-1})\propto w_{0:k}^{(\gamma_{0:k})}\label{eq:jointpi}%
\end{equation}
where $w_{0:k}^{(\gamma_{0:k})}$ is given by (\ref{eq:MSGLMBCannonical1}).

We present two techniques for sampling from (\ref{eq:jointpi}). The first is
based on sampling from the factors (\ref{eq:conditionals0}), i.e., $\gamma
_{j}\sim\pi^{(j)}(\cdot|\gamma_{0:j-1})$, for $j=1:k$, using the Gibbs sampler
of \cite{VVH17}. The alternative is a full Gibbs sampler with
(\ref{eq:jointpi}) as the stationary distribution.

\subsection{Sampling from the Factors}

Sampling from (\ref{eq:conditionals0}) using the Gibbs sampler
\cite{Geman_Gibbs84}, \cite{Cassella_Gibbs92} involves constructing a Markov
chain where a new state $\gamma_{j}^{\prime}$ is generated from state
$\gamma_{j}$ by sampling the values of $\gamma_{j}^{\prime}(\ell_{n})$,
$\ell_{n}\in\{\ell_{1},...,\ell_{|\mathbb{L}_{j}|}\}\triangleq\mathbb{L}_{j}$
from the following distribution
\[
\pi_{n}^{(j)}(\alpha|\gamma_{j}^{\prime}(\ell_{1:n-1}),\gamma_{j}%
(\ell_{n+1:|\mathbb{L}_{j}|}),\gamma_{0:j-1})\triangleq\pi^{(j)}(\gamma
_{j}(\ell_{n})\text{=}\alpha|\gamma_{j}^{\prime}(\ell_{1:n-1}),\gamma_{j}%
(\ell_{n+1:|\mathbb{L}_{j}|}),\gamma_{0:j-1},)
\]
where%
\begin{align*}
\pi^{(j)}(\gamma_{j}(\ell_{n})|\gamma_{j}(\ell_{\bar{n}}),\gamma_{0:j-1})  &
\propto\pi^{(j)}(\gamma_{j}|\gamma_{0:j-1}).\\
\gamma_{j}(\ell_{u:v})  &  \triangleq\lbrack\gamma_{j}(\ell_{u}),...,\gamma
_{j}(\ell_{v})]\\
\gamma_{j}(\ell_{\bar{n}})  &  \triangleq\lbrack\gamma_{j}(\ell_{1:n-1}%
),\gamma_{j}(\ell_{n+1:|\mathbb{L}_{j}|})]
\end{align*}

For a valid $\gamma_{j}$, i.e. $\pi^{(j)}(\gamma_{j}|\gamma_{0:j-1})>0$, it is
necessary that $1_{\mathcal{F}(\mathbb{B}_{i}\uplus\mathcal{L(}\gamma_{i-1}%
))}(\mathcal{L}(\gamma_{j}))=1$ in (\ref{eq:conditionals0}), i.e.
$\mathcal{L}(\gamma_{j})\subseteq\mathbb{B}_{j}\uplus\mathcal{L(}\gamma
_{j-1})$. This amounts to disregarding any $\gamma_{j}$ that takes on a
non-negative value outside $\mathbb{B}_{j}\mathbf{\uplus}\mathcal{L(}%
\gamma_{j-1})$, and only consider those that take on -1 everywhere outside of
$\mathbb{B}_{j}\mathbf{\uplus}\mathcal{L(}\gamma_{j-1})$, in which case
\[
\pi^{(j)}(\gamma_{j}(\ell_{n})|\gamma_{j}(\ell_{\bar{n}}),\gamma
_{0:j-1})\propto1_{\Gamma_{j}}(\gamma_{j})\prod\limits_{\ell\in\mathbb{B}%
_{i}\uplus\mathcal{L(}\gamma_{i-1})}\omega_{j|j-1}^{(\gamma_{0:j}(\ell))}.
\]
for $\ell_{n}\in\{\ell_{1},...,\ell_{|\mathbb{B}_{j}\uplus\mathcal{L(}%
\gamma_{j-1})|}\}\triangleq\mathbb{B}_{j}\uplus\mathcal{L(}\gamma_{j-1})$.
Further, applying Proposition 3 of \cite{VVH17} gives:%
\[
\pi^{(j)}(\gamma_{j}(\ell_{n})|\gamma_{j}(\ell_{\bar{n}}),\gamma
_{0:j-1})\propto\left\{
\begin{array}
[c]{ll}%
\omega_{j|j-1}^{(\gamma_{0:j}(\ell_{n}))}, & \gamma_{j}(\ell_{n})\leq0\\
\omega_{j|j-1}^{(\gamma_{0:j}(\ell_{n}))}(1-1_{\gamma_{j}(\ell_{\bar{n}}%
)}(\gamma_{j}(\ell_{n}))), & \gamma_{j}(\ell_{n})>0
\end{array}
\right.  .
\]

Hence, to generate $\gamma_{j}^{\prime}$ from a valid $\gamma_{j}$, we set
$\gamma_{j}^{\prime}(\ell)=-1$ for all $\ell\in\mathbb{L}_{j}-\mathbb{B}%
_{j}\mathbf{\uplus}\mathcal{L(}\gamma_{j-1})$ and sample $\gamma_{j}^{\prime
}(\ell_{n})$ for $\ell_{n}\in\{\ell_{1},...,\ell_{|\mathbb{B}_{j}%
\uplus\mathcal{L(}\gamma_{j-1})|}\}$ from $\pi_{n}^{(j)}(\cdot|\gamma
_{j}^{\prime}(\ell_{1:n-1}),\gamma_{j}(\ell_{n+1:|\mathbb{L}_{j}|}%
),\gamma_{0:j-1})$. Note that in implementation, we only need the values of
$\gamma_{j}$ on $\mathbb{B}_{j}\uplus\mathcal{L(}\gamma_{j-1})$. The pseudo
code for sampling from (\ref{eq:jointpi}) by sampling from the factors is
given in Algorithm 1.

\vspace{8pt}

\hrule

\vspace{8pt}

\vspace{5pt}

\textbf{Algorithm 1 }

\begin{itemize}
\item \textsf{{\footnotesize {input:}}}$\ T_{1:k};$
\textsf{{\footnotesize {output: }}}$\gamma_{1:k}$
\end{itemize}

\vspace{2pt}

\hrule

\vspace{2pt}

$\mathcal{L(}\gamma_{0})=\{\};$

\textsf{{\footnotesize {for }}}$j=1:k$

$\quad P:=|\mathbb{B}_{j}\uplus\mathcal{L(}\gamma_{j-1})|;M:=|Z_{j}%
|;\ c:=[-1$:$M];\ \gamma^{(1)}=${ }\textsf{{\footnotesize {zeros(}}}%
$P$\textsf{{\footnotesize {); or some other methods}}}

\quad\textsf{{\footnotesize {for}}\ }$t=2:T_{j}$

\quad\quad$\gamma^{(t)}:=[$ $];$

\quad\quad\textsf{{\footnotesize {for }}}$n=1:P$

\quad\quad\quad\textsf{{\footnotesize {for }}}$\alpha=-1:M$

\quad\quad\quad\quad$\eta_{n}(\alpha):=\pi_{n}^{(j)}(\alpha|\gamma_{j}%
^{(t)}(\ell_{1:n-1}),\gamma_{j}^{(t-1)}(\ell_{n+1:P}),\gamma_{0:j-1});$

\quad\quad\quad\textsf{{\footnotesize {end}}}

\quad\quad\quad$\gamma_{n}^{(t)}\sim\mathsf{Categorical}(c,\eta_{n});$%
\quad$\gamma^{(t)}:=[\gamma^{(t)},\gamma_{n}^{(t)}];$

\quad\quad\textsf{{\footnotesize {end}}}

\quad\textsf{{\footnotesize {end}}}

\quad$\gamma_{j}=\gamma^{(t)};$

\textsf{{\footnotesize {end}}}

\vspace{5pt}

\hrule

\bigskip

This approach ensures that the sample $\gamma_{1:k}$ is a valid association
history. However, we have to run each Gibbs sampler sufficiently long to
ensure that sample $\gamma_{j}$ is close enough to being a sample from
$\pi^{(j)}(\cdot|\gamma_{0:j-1})$ so that $\gamma_{1:k}$ is close enough to
being a sample from (\ref{eq:jointpi}). Nonetheless, sampling from the factors
can be use to generate a good starting point for the full Gibbs sampler.

\subsection{Gibbs Sampling}

Sampling from (\ref{eq:jointpi}) using the Gibbs sampler \cite{Geman_Gibbs84},
\cite{Cassella_Gibbs92} involves constructing a Markov chain where a new state
$\gamma_{1:k}^{\prime}$ is generated from state $\gamma_{1:k}$ by sampling the
values of the components $\gamma_{j}^{\prime}(\ell_{n})$, $j=1:k$, $\ell
_{n}\in\{\ell_{1},...,\ell_{|\mathbb{L}_{j}|}\}=\mathbb{L}_{j}$ according to
the conditional distribution%
\[
\pi(\gamma_{j}^{\prime}(\ell_{n})|\gamma_{0:j-1}^{\prime},\gamma_{j}^{\prime
}(\ell_{1:n-1}),\gamma_{j}(\ell_{n+1:|\mathbb{L}_{j}|}),\gamma_{j+1:k}%
)\propto\pi(\gamma_{0:j-1}^{\prime},\gamma_{j}^{\prime}(\ell_{1:n}),\gamma
_{j}(\ell_{n+1:|\mathbb{L}_{j}|}),\gamma_{j+1:k}).
\]
Observe from (\ref{eq:conditionals0}) and (\ref{eq:jointpi}) that for a valid
$\gamma_{1:k}$, i.e. $\pi(\gamma_{1:k})>0$, it is necessary that
$1_{\Gamma_{i}}(\gamma_{i})=1$ (i.e. $\gamma_{i}$ is positive 1-1), and
$1_{\mathcal{F}(\mathbb{B}_{i}\uplus\mathcal{L(}\gamma_{i-1}))}(\mathcal{L}%
(\gamma_{i}))=1$ (i.e. dead labels at $i-1$ cannot be live at $i$, or
equivalently, a live label at $i$ cannot be dead at $i-1$) for $i=1:k$. Thus,
in addition to being positive 1-1, consecutive elements of a valid
$\gamma_{1:k}$ must be such that dead labels remains dead at the next time.
Closed form expressions for the conditionals are given in the following
Proposition (see Appendix \ref{sec:AppendixConditionals} for proof).

\begin{proposition}
\label{conditionals}Suppose $\gamma_{j}:\mathbb{L}_{j}\rightarrow\{-1$%
:$|Z_{j}|\}$, $j\in\{1:k\}$, is an association map of a valid association
history $\gamma_{1:k}$. Then, for $\ell_{n}$ $\in$ $\{\ell_{1},...,\ell
_{|\mathbb{B}_{j}\uplus\mathcal{L(}\gamma_{j-1})|}\}\triangleq\mathbb{B}%
_{j}\mathbf{\uplus}\mathcal{L(}\gamma_{j-1})$,%
\begin{equation}
\pi(\gamma_{j}(\ell_{n})|\gamma_{j}(\ell_{\bar{n}}),\gamma_{\bar{j}}%
)\propto\left\{
\begin{array}
[c]{ll}%
\prod\limits_{i=j}^{k}\omega_{i|i-1}^{(\gamma_{0:i}(\ell_{n}))}\delta
_{\gamma_{\min\{j+1,k\}}(\ell_{n})}[\gamma_{j}(\ell_{n})], & \gamma_{j}%
(\ell_{n})<0\\
\prod\limits_{i=j}^{k}\omega_{i|i-1}^{(\gamma_{0:i}(\ell_{n}))}, & \gamma
_{j}(\ell_{n})=0\\
\prod\limits_{i=j}^{k}\omega_{i|i-1}^{(\gamma_{0:i}(\ell_{n}))}(1-1_{\gamma
_{j}(\ell_{\bar{n}})}(\gamma_{j}(\ell_{n}))), & \gamma_{j}(\ell_{n})>0
\end{array}
\right.  \label{eq:FullGibbsCond}%
\end{equation}
and for $\ell_{n}$ $\in$ $\{\ell_{|\mathbb{B}_{j}\uplus\mathcal{L(}%
\gamma_{j-1})|+1},...,\ell_{|\mathbb{L}_{j}|}\}\triangleq\mathbb{L}%
_{j}-\mathbb{B}_{j}\mathbf{\uplus}\mathcal{L(}\gamma_{j-1})$%
\begin{equation}
\pi(\gamma_{j}(\ell_{n})|\gamma_{j}(\ell_{\bar{n}}),\gamma_{\bar{j}}%
)=\delta_{-1}[\gamma_{j}(\ell_{n})]\delta_{\gamma_{\min\{j+1,k\}}(\ell_{n}%
)}[\gamma_{j}(\ell_{n})]. \label{eq:FullGibbsCond1}%
\end{equation}

\end{proposition}

To generate $\gamma_{j}^{\prime}$, from a valid $\gamma_{1:k}$, we sample
$\gamma_{j}^{\prime}(\ell_{n})$, $\ell_{n}\in\{\ell_{1},...,\ell
_{|\mathbb{B}_{j}\uplus\mathcal{L(}\gamma_{j-1})|}\}$ from $\pi(\cdot
|\gamma_{0:j-1}^{\prime},\gamma_{j}^{\prime}(\ell_{1:n-1}),\gamma_{j}%
(\ell_{n+1:|\mathbb{L}_{j}|}),\gamma_{j+1:k})$ as given by
(\ref{eq:FullGibbsCond}), and set $\gamma_{j}^{\prime}(\ell_{n})=-1$ for the
remaining $\ell_{n}$. This last step is omitted in actual implementation and
it is understood that $\gamma_{j}^{\prime}$ is negative outside of $\{\ell
_{1},...,\ell_{|\mathbb{B}_{j}\uplus\mathcal{L(}\gamma_{j-1})|}\}$. The Gibbs
sampler with stationary distribution (\ref{eq:jointpi}) can be constructed as follows:

\bigskip

\hrule

\vspace{8pt}

\vspace{5pt}

\textbf{Algorithm 2: Full Gibbs }

\begin{itemize}
\item \textsf{{\footnotesize {input: }}}$\gamma_{1:k}^{(1)}{\ }$%
\textsf{{\footnotesize {(use Algorithm 1) }}}$T,$

\item \textsf{{\footnotesize {output: }}}$\gamma_{1:k}^{(1)},...,\gamma
_{1:k}^{(T)}$
\end{itemize}

\vspace{2pt}

\hrule

\vspace{4pt}

\textsf{{\footnotesize {for}}\ }$t=1:T$

\quad$\gamma_{0}^{(t)}=\{\}$

\quad\textsf{{\footnotesize {for }}}$j=1:k$

\quad\quad$P:=|\mathbb{B}_{j}\uplus\mathcal{L(}\gamma_{j-1}^{(t)}%
)|;\ M:=|Z_{j}|;\ c:=[-1$:$M];$

\quad\quad\textsf{{\footnotesize {for }}}$n=1:P$

\quad\quad\quad\textsf{{\footnotesize {for }}}$\alpha=-1:M$

\quad\quad\quad\quad$\eta_{n}(\alpha):=\pi(\alpha|\gamma_{0:j-1}^{(t)}%
,\gamma_{j}^{(t)}(\ell_{1:n-1}),\gamma_{j}^{(t-1)}(\ell_{n+1:P}),\gamma
_{j+1:k}^{(t-1)});$

\quad\quad\quad\textsf{{\footnotesize {end}}}

\quad\quad\quad$\gamma_{j}^{(t)}(\ell_{n})\sim\mathsf{Categorical}(c,\eta
_{n});$ $\gamma_{j}^{(t)}:=[\gamma_{j}^{(t)};\gamma_{j}^{(t)}(\ell_{n})];$

\quad\quad\textsf{{\footnotesize {end}}}

\quad\textsf{{\footnotesize {end}}}

\textsf{{\footnotesize {end}}}

\vspace{5pt}

\hrule

\bigskip

Starting with a valid association history, it follows from Proposition
\ref{conditionals} that all iterates of the Gibbs sampler (described above)
are also valid association histories.

\begin{proposition}
\label{convergence} Starting from any valid initial state, the Gibbs sampler
defined by the family of conditionals (\ref{eq:FullGibbsCond}) converges to
the target distribution (\ref{eq:jointpi}) at an exponential rate. More
concisely, let $\pi^{j}$ denote the $j$th power of the transition kernel,
then
\[
\max_{\gamma_{1:k},\gamma_{1:k}^{\prime}\in{\Gamma}_{k}}(|\pi^{j}(\gamma
_{1:k}^{\prime}|\gamma_{1:k})-\pi(\gamma_{1:k}^{\prime})|)\leq(1-2\beta
)^{\left\lfloor \frac{j}{h}\right\rfloor },
\]
where, $h=k+1,$ $\beta\triangleq\min_{\gamma_{1:k},\gamma_{1:k}^{\prime}%
\in{\Gamma}_{k}}\pi^{h}(\gamma_{1:k}^{\prime}|\gamma_{1:k})>0$ is the least
likely $h$-step transition probability.
\end{proposition}

The proof follows along the same line as Proposition 4 of \cite{VVH17}, with
the 2-step transition probability replaced by the ($k+1$)-step transition
probability. Instead of going from one arbitrary state of the chain to another
via the all-zeros state in 2 steps as in \cite{VVH17}, in this case we go to
the all-negative state (consists of all -1) in $k$ steps or less, and from
this state to the other state in one additional step.

Similar to Gibbs sampling for the 2D assignment problem, the proposed Gibbs
sampler has a relatively short burn-in period. For the purpose of
approximating the multi-scan GLMB posterior density, it is not necessary to
wait for samples from the stationary distribution since each distinct sample
constitutes one term in the approximant, and reduces the $L_{1}$ approximation
error by an amount proportional to its weight.

Recall that one of the proposed estimator is based on the most significant
$\gamma_{1:k}$. The full Gibbs sampler above can be used in the simulated
annealing setting to find the best $\gamma_{1:k}$ more efficiently.

\subsection{Computing Multi-Scan GLMB Components}

This subsection details the computations the multi-scan GLMB
(\ref{eq:MSGLMBCannonical0}) parameters for linear Gaussian multi-object
models, i.e.%
\begin{align*}
\psi_{k,\{z_{1},...,z_{m}\}}^{(j)}(x,\ell)  &  =\left\{
\begin{array}
[c]{ll}%
\frac{P_{D,k}^{(\ell)}\mathcal{N}(z_{j};Hx,R_{k})}{\kappa_{k}(z_{j})}, &
\text{if }j>0\\
Q_{D,k}^{(\ell)} & \text{if }j=0
\end{array}
\right. \\
P_{S,k-1}(\varsigma,\ell)  &  =P_{S,k-1}^{(\ell)}\\
f_{S.k|k-1}(x|\varsigma,\ell)  &  =\mathcal{N}(x;F_{k|k-1}\varsigma,Q_{k})\\
f_{B,k}(x,\ell)  &  =\mathcal{N}(x;m_{B,k}^{(\ell)},Q_{B,k}^{(\ell)})
\end{align*}
where the single-object state $x$ is a $d$-dimensional vector, $\mathcal{N}%
(\cdot;m,P)$ denotes a Gaussian density with mean $m$ and covariance $P$.

It follows from (\ref{eq:MSGLMBCannonical3}), (\ref{eq:MSGLMBCannonical4})
that the single object densities $p_{0:k}^{(\gamma_{0:k})}(\cdot,\ell)$ are
Gaussians. Further, $p_{0:k}^{(\gamma_{0:k})}(\cdot,\ell)$ (and $\omega
_{k|k-1}^{(\gamma_{0:k}(\ell))}$ for the Gibbs sampler) can be computed
recursively using the following standard results on joint and conditional
Gaussians%
\begin{align*}
\mathcal{N}(z;Hx,R)\mathcal{N}(x;m,P)  &  =q(z;H,R,m,P)\mathcal{N}\left(
x;\mu(z,H,R,m,P),\Sigma(H,R,P)\right) \\
&  =\mathcal{N}\left(  \left[  x;z\right]  ,\hat{\mu}(H,m),\hat{\Sigma
}(H,R,P)\right)
\end{align*}
where%
\begin{align*}
q(z;H,R,m,P)  &  \triangleq\mathcal{N}\left(  z;Hm,R+HPH^{T}\right) \\
\mu(z,H,R,m,P)  &  \triangleq m+PH^{T}(HPH^{T}+R)^{-1}(z-Hm)\\
\Sigma(H,R,P)  &  \triangleq P-PH^{T}(HPH^{T}+R)^{-1}HP\\
\hat{\mu}(H,m)  &  \triangleq\left[
\begin{array}
[c]{c}%
m\\
Hm
\end{array}
\right]  ,\text{ }\hat{\Sigma}(H,R,P)\triangleq\left[
\begin{array}
[c]{cc}%
P & PH^{T}\\
HP & R+HPH^{T}%
\end{array}
\right]
\end{align*}

If $\ell\notin\mathcal{L(}\gamma_{k})$, then for $\ell\in\mathbb{B}%
_{k}-\mathcal{L(}\gamma_{k})$, i.e. label $\ell$ is unborn, $\omega
_{k|k-1}^{(\gamma_{0:k}(\ell))}=Q_{B,k}^{(\ell)}$, and for $\ell
\in\mathcal{L(}\gamma_{k-1})-\mathcal{L(}\gamma_{k})$, i.e. label $\ell$ died,
$p_{0:k}^{(\gamma_{0:k})}(\cdot,\ell)=p_{0:k-1}^{(\gamma_{0:k-1})}(\cdot
;\ell)$ and $\omega_{k|k-1}^{(\gamma_{0:k}(\ell))}=Q_{S,k-1}^{(\ell)}$.

Otherwise $\ell\in\mathcal{L(}\gamma_{k})$, and we have the following cases.

Case: $s(\ell)=k$,%
\begin{align*}
p_{0:k}^{(\gamma_{0:k})}(x_{s(\ell):t(\ell)};\ell) &  =\mathcal{N}%
(x_{k};m_{B,k}^{(\ell)},Q_{B,k}^{(\ell)})\\
\omega_{k|k-1}^{(\gamma_{0:k}(\ell))} &  =\bar{\Lambda}_{B,k}^{(\gamma
_{k}(\ell))}(\ell)=P_{B,k}^{(\ell)}Q_{D,k}^{(\ell)}%
\end{align*}
Further if $\gamma_{k}(\ell)>0$%
\begin{align*}
p_{0:k}^{(\gamma_{0:k})}(x_{s(\ell):t(\ell)}^{(\ell)};\ell) &  \propto
\mathcal{N}(z_{\gamma_{k}(\ell)};H_{k}x_{k},R_{k})\mathcal{N}(x_{k}%
;m_{B,k}^{(\ell)},Q_{B,k}^{(\ell)})\\
&  =\mathcal{N}(x_{k};m^{(\gamma_{0:k})}(\ell),P^{(\gamma_{0:k})}%
(\ell))q(z_{\gamma_{k}(\ell)};H_{k},R_{k},m_{B,k}^{(\ell)},Q_{B,k}^{(\ell)})\\
m^{(\gamma_{0:k})}(\ell) &  =\mu(z_{\gamma_{k}(\ell)},H_{k},R_{k}%
,m_{B,k}^{(\ell)},Q_{B,k}^{(\ell)})\\
P^{(\gamma_{0:k})}(\ell) &  =\Sigma(H_{k},R_{k},Q_{B,k}^{(\ell)})\\
\omega_{k|k-1}^{(\gamma_{0:k}(\ell))} &  =\bar{\Lambda}_{B,k}^{(\gamma
_{k}(\ell))}(\ell)=P_{B,k}^{(\ell)}P_{D,k}^{(\ell)}\frac{q(z_{\gamma_{k}%
(\ell)};H_{k},R_{k},m_{B,k}^{(\ell)},Q_{B,k}^{(\ell)})}{\kappa_{k}%
(z_{\gamma_{k}(\ell)})}%
\end{align*}

Case: $s(\ell)<k$, then $p_{0:k-1}^{(\gamma_{0:k-1})}(\cdot;\ell)$ has the
form%
\[
p_{0:k-1}^{(\gamma_{0:k-1})}(x_{s(\ell):k-1};\ell)=\mathcal{N}(x_{s(\ell
):k-1},m^{(\gamma_{0:k-1})}(\ell),P^{(\gamma_{0:k-1})}(\ell))
\]
and
\begin{align*}
p_{0:k}^{(\gamma_{0:k})}(x_{s(\ell):k}^{(\ell)};\ell) &  \propto
\mathcal{N}(x_{k};F_{k}^{(\ell)}x_{k-1},Q_{k}^{(\ell)})\mathcal{N}%
(x_{s(\ell):k-1},m^{(\gamma_{0:k-1})}(\ell),P^{(\gamma_{0:k-1})}(\ell))\\
&  =\mathcal{N}(x_{s(\ell):k},\hat{m}^{(\gamma_{0:k})}(\ell),\hat{P}%
^{(\gamma_{0:k})}(\ell))\\
\hat{m}^{(\gamma_{0:k})}(\ell) &  =\hat{\mu}(F_{k|k-1}\Pi_{k-1}^{(\ell
)},m^{(\gamma_{0:k-1})}(\ell))\\
\hat{P}^{(\gamma_{0:k})}(\ell) &  =\hat{\Sigma}(F_{k|k-1}\Pi_{k-1}^{(\ell
)},Q_{k},P^{(\gamma_{0:k-1})}(\ell))\\
\Pi_{j}^{(\ell)} &  =\left[
\begin{array}
[c]{cc}%
0_{d,(j-s(\ell))d}, & I_{d,d}%
\end{array}
\right]  \\
\omega_{k|k-1}^{(\gamma_{0:k}(\ell))} &  =\bar{\Lambda}_{S.k|k-1}%
^{(\gamma_{0:k-1},\gamma_{k}(\ell))}(\ell)=\int\Lambda_{S,k}^{(\gamma_{k}%
(\ell))}(x_{k},\ell|x_{k-1})p_{s(\ell):k-1}^{(\gamma_{0:k-1})}(x_{s(\ell
):k-1};\ell)dx_{s(\ell):k}\\
&  =P_{S,k-1}^{(\ell)}Q_{D,k}^{(\ell)}\int\mathcal{N}(x_{s(\ell):k},\hat
{m}^{(\gamma_{0:k})}(\ell),\hat{P}^{(\gamma_{0:k})}(\ell))dx_{s(\ell):k}\\
&  =P_{S,k-1}^{(\ell)}Q_{D,k}^{(\ell)}%
\end{align*}
Note that the corresponding prediction density to time $k$ is given by the
marginal $\mathcal{N}(x_{k},\hat{m}_{k}^{(\gamma_{0:k})}(\ell),\hat{P}%
_{k}^{(\gamma_{0:k})}(\ell))$, where $\hat{m}_{k}^{(\gamma_{0:k})}%
(\ell)\triangleq\Pi_{k}^{(\ell)}\hat{m}^{(\gamma_{0:k})}(\ell)$, $\hat{P}%
_{k}^{(\gamma_{0:k})}(\ell)\triangleq\Pi_{k}^{(\ell)}\hat{P}_{k}%
^{(\gamma_{0:k})}(\ell)(\Pi_{k}^{(\ell)})^{T}$. Further if $\gamma_{k}%
(\ell)>0$%
\begin{align*}
p_{0:k}^{(\gamma_{0:k})}(x_{s(\ell):k}^{(\ell)};\ell) &  \propto
\mathcal{N}(z_{\gamma_{k}(\ell)},H_{k}x_{k},R_{k})\mathcal{N}(x_{s(\ell
):k},\hat{m}^{(\gamma_{0:k})}(\ell),\hat{P}^{(\gamma_{0:k})}(\ell))\\
&  =\mathcal{N}(x_{s(\ell):k},m^{(\gamma_{0:k})}(\ell),P^{(\gamma_{0:k})}%
(\ell))q(z_{\gamma_{k}(\ell)};H_{k}\Pi_{k}^{(\ell)},R_{k},\hat{m}%
^{(\gamma_{0:k})}(\ell),\hat{P}^{(\gamma_{0:k})}(\ell))\\
&  =\mathcal{N}(x_{s(\ell):k},m^{(\gamma_{0:k})}(\ell),P^{(\gamma_{0:k})}%
(\ell))q(z_{\gamma_{k}(\ell)};H_{k},R_{k},\hat{m}_{k}^{(\gamma_{0:k})}%
(\ell),\hat{P}_{k}^{(\gamma_{0:k})}(\ell))\\
m^{(\gamma_{0:k})}(\ell) &  =\mu(z_{\gamma_{k}(\ell)},H_{k}\Pi_{k}^{(\ell
)},R_{k},\hat{m}^{(\gamma_{0:k})}(\ell),\hat{P}^{(\gamma_{0:k})}(\ell))\\
P^{(\gamma_{0:k})}(\ell) &  =\Sigma(H_{k}\Pi_{k}^{(\ell)},R_{k},\hat
{P}^{(\gamma_{0:k})}(\ell))\\
\omega_{k|k-1}^{(\gamma_{0:k}(\ell))} &  =\bar{\Lambda}_{S.k|k-1}%
^{(\gamma_{0:k-1},\gamma_{k}(\ell))}(\ell)=\int\Lambda_{S,k}^{(\gamma_{k}%
(\ell))}(x_{k},\ell|x_{k-1})p_{s(\ell):k-1}^{(\gamma_{0:k-1})}(x_{s(\ell
):k-1};\ell)dx_{s(\ell):k}\\
&  =P_{S,k-1}^{(\ell)}P_{D,k}^{(\ell)}\frac{q(z_{\gamma_{k}(\ell)};H_{k}%
,R_{k},\hat{m}_{k}^{(\gamma_{0:k})}(\ell),\hat{P}_{k}^{(\gamma_{0:k})}(\ell
))}{\kappa_{k}(z_{\gamma_{k}(\ell)})}%
\end{align*}

\section{Numerical Experiments$\label{sec:Numericals}$}

This section presents a preliminary numerical study demonstrating the
improvements in tracking performance of the proposed multi-scan GLMB tracker
over the single-scan counterpart \cite{VVH17}. For this purpose we use the
linear Gaussian scenario shown in Section IV(A) of \cite{VVH17}. This scenario
involves an unknown and time varying number of targets (up to 10 in total)
over 100 time steps with births, deaths and crossings. Individual object
kinematics are described by a 4D state vector\ of position and velocity that
follows a constant velocity model with sampling period of\ $1s$, and process
noise standard deviation $\sigma_{\nu}=5m/s^{2}$. The survival probability
$P_{S}=0.99$, and the birth model is an LMB with parameters $\{r_{B,k}%
(\ell_{i}),p_{B,k}(\ell_{i})\}_{i=1}^{3}$, where $\ell_{i}=(k,i)$,
$r_{B,k}(\ell_{i})=0.04$, and $p_{B}(x,\ell_{i})=\mathcal{N}(x;m_{B}%
^{(i)},P_{B})$ with%
\[%
\begin{array}
[c]{ll}%
m_{B}^{(1)}=[0,0,100,0]^{T}, & \!\!m_{B}^{(2)}=[-100,0,-100,0]^{T},\\
m_{B}^{(3)}=[100,0,-100,0]^{T}, & \!\!P_{B}^{\text{ \ \ \ }}=\mathrm{diag}%
([10,10,10,10]^{T})^{2}.
\end{array}
\]

Observations are 2D position vectors on the region $[-1000,1000]m\times
\lbrack-1000,1000]m$ with noise standard deviation $\sigma_{\varepsilon}=10m$.
Clutter is modeled as a Poisson RFS with a uniform intensity of $\lambda
_{c}=1.65\times10^{-5}~m^{-2}$ on the observation region (i.e. an average of
66 false alarms per scan). The detection probability $P_{D}=0.77$, which is
lower than the original value of 0.88 in \cite{VVH17}.

\begin{figure}[h]
\begin{center}
\resizebox{80mm}{!}{\includegraphics{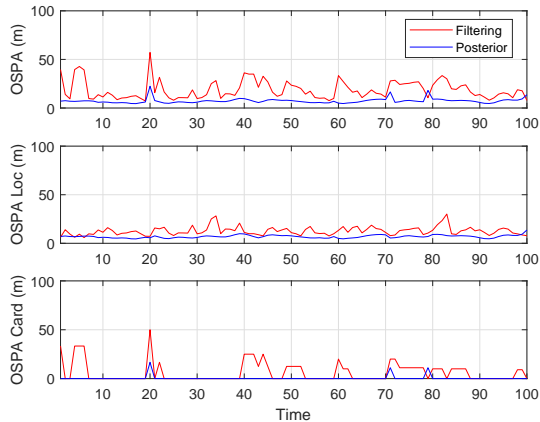}}
\end{center}
\caption{OSPA error along with localization and cardinality components:
evaluation of the instantaneous filtering performance at each individual scan}%
\label{fig:ospa}%
\end{figure}

\begin{figure}[h]
\begin{center}
\resizebox{80mm}{!}{\includegraphics{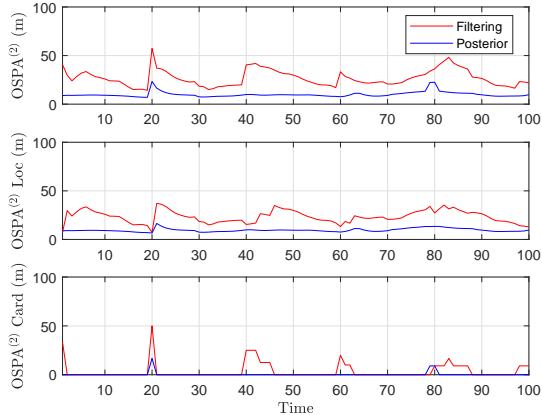}}
\end{center}
\caption{OSPA$^{(2)}$ error along with localization and cardinality
components: evaluation of the tracking performance over a lagging 10 scan
window}%
\label{fig:ospa2}%
\end{figure}

The standard GLMB tracker \cite{VVH17} is run with a maximum $10000$
components. The proposed multi-scan GLMB tracker is run for 100 iterations of
the multi-scan Gibbs sampler. The multi-target \emph{filtering} errors are
evaluated using the OSPA\ metric with parameters $c=100m$ and $p=1$ as shown
in Figure 1. In addition the multi-target \emph{tracking} errors are evaluated
using the OSPA$^{(2)}$ metric \cite{BVVarxivLST18} with the same parameters
and a window length of 10 time steps. Observe from Figure 1 that there is a
significant improvement in the \emph{instantaneous} localization and
cardinality errors. Furthermore the results shown in Figure 2 confirm a
significant improvement in the \emph{trajectory} errors over the entire
scenario duration.

\section{Conclusions$\label{sec:Conclusions}$}

By introducing the multi-scan GLMB model, we extended the GLMB filtering
recursion to propagate the labeled multi-object posterior density, i.e.
multi-object smoothing-while-filtering. Conceptually the multi-scan GLMB
recursion is more intuitive, but numerically it is far more challenging than
the (single-scan) GLMB recursion. We showed that computing the multi-scan GLMB
posterior with minimal $L_{1}$-error (from its exact value) requires solving a
multi-dimensional assignment problem with very high dimensions. Further, we
developed an efficient and highly parallelizable algorithm for solving such
multi-dimensional assignment problems using Gibbs sampling, and subsequently a
novel multi-object smoothing-while-filtering algorithm. Numerical multi-object
tracking examples demonstrated that the proposed algorithm significantly
improves tracking performance as well as eliminating track fragmentation, a
problem often found in multi-object filters. This is possible\ because the
labeled multi-object posterior contains all information on the entire history
of the underlying trajectories.

\section{Appendix$\label{sec:Appendix}$}

\subsection{Properties of Multi-scan
Exponentials$\label{sec:AppendixProperties}$}

To present relevant properties of multi-scan exponentials, we introduce some
useful partitionings for the labels of the multi-object state sequence
$\mathbf{X}_{j:k}$. Given a time $i$ in $\{j:k\}$, a label $\ell\in\cup
_{r=j}^{k}\mathcal{L}(\mathbf{X}_{r}\mathbf{)}$ is alive at $i$ iff $\ell
\in\mathcal{L}(\mathbf{X}_{i}\mathbf{)}$, terminates at $t(\ell)<i$ (before
$i$) iff $\ell\in$ $\mathcal{L}(\mathbf{X}_{t}\mathbf{)}-\mathcal{L(}%
\mathbf{X}_{t+1})$, and born at time $s(\ell)>i$ (after $i$) iff $\ell\in$
$\mathcal{L}(\mathbf{X}_{s}\mathbf{)}\cap\mathbb{B}_{s}$. The set of labels in
$\mathbf{X}_{j:k}$ can be partitioned into labels terminated before $i$, live
labels at $i$, and labels born after $i$, i.e.
\begin{equation}%
%TCIMACRO{\tbigcup _{r=j}^{k}}%
%BeginExpansion
{\textstyle\bigcup_{r=j}^{k}}
%EndExpansion
\mathcal{L}(\mathbf{X}_{r}\mathbf{)}=\overleftarrow{\mathcal{L(}\mathbf{X}%
_{i})}\mathbf{\uplus}\mathcal{L(}\mathbf{X}_{i})\mathbf{\uplus}\overrightarrow
{\mathcal{L}(\mathbf{X}_{i}\mathbf{)}} \label{eq:labelpartition0}%
\end{equation}
where
\begin{align}
\overleftarrow{\mathcal{L(}\mathbf{X}_{i})}  &  \triangleq\left\{  \ell\in
\cup_{r=j}^{k}\mathcal{L}(\mathbf{X}_{r}\mathbf{):}t(\ell)<i\right\}  =%
%TCIMACRO{\tbigcup _{r=j}^{i}}%
%BeginExpansion
{\textstyle\bigcup_{r=j}^{i}}
%EndExpansion
\mathcal{L}(\mathbf{X}_{r}\mathbf{)-}\mathcal{L}(\mathbf{X}_{i}\mathbf{)}=%
%TCIMACRO{\tbiguplus \limits_{t=j}^{i-1}}%
%BeginExpansion
{\textstyle\biguplus\limits_{t=j}^{i-1}}
%EndExpansion
(\mathcal{L}(\mathbf{X}_{t}\mathbf{)}-\mathcal{L(}\mathbf{X}_{t+1}%
))\label{eq:labelpartition1}\\
\overrightarrow{\mathcal{L}(\mathbf{X}_{i}\mathbf{)}}  &  =\left\{  \ell
\in\cup_{r=j}^{k}\mathcal{L}(\mathbf{X}_{r}\mathbf{):}s(\ell)>i\right\}  =%
%TCIMACRO{\tbigcup _{r=i}^{k}}%
%BeginExpansion
{\textstyle\bigcup_{r=i}^{k}}
%EndExpansion
\mathcal{L}(\mathbf{X}_{r}\mathbf{)-}\mathcal{L}(\mathbf{X}_{i}\mathbf{)=}%
%TCIMACRO{\tbiguplus \limits_{s=i+1}^{k}}%
%BeginExpansion
{\textstyle\biguplus\limits_{s=i+1}^{k}}
%EndExpansion
\mathcal{L}(\mathbf{X}_{s}\mathbf{)}\cap\mathbb{B}_{s}.
\label{eq:labelpartition2}%
\end{align}

When $i=k$, $\overrightarrow{\mathcal{L}(\mathbf{X}_{k}\mathbf{)}}=\emptyset$
and the set of labels in $\mathbf{X}_{j:k}$ can be partitioned into labels
terminated before $k$ and live labels at $k$, i.e. (\ref{eq:labelpartition1})
becomes
\begin{equation}%
%TCIMACRO{\tbigcup _{r=j}^{k}}%
%BeginExpansion
{\textstyle\bigcup_{r=j}^{k}}
%EndExpansion
\mathcal{L}(\mathbf{X}_{r}\mathbf{)=}\overleftarrow{\mathcal{L(}\mathbf{X}%
_{k})}\mathbf{\uplus}\mathcal{L}(\mathbf{X}_{k}). \label{eq:labelpartition3}%
\end{equation}
In addition, if $j=k-1$, then $\overleftarrow{\mathcal{L(}\mathbf{X}_{k}%
)}=\mathcal{L}(\mathbf{X}_{k-1}\mathbf{)}-\mathcal{L(}\mathbf{X}_{k})$, and
using the decomposition $\mathcal{L}(\mathbf{X}_{k})=\mathbf{(}\mathcal{L(}%
\mathbf{X}_{k-1})\cap\mathcal{L(}\mathbf{X}_{k}))\mathbf{\uplus(}%
\mathbb{B}_{k}\cap\mathcal{L(}\mathbf{X}_{k}))$, the set of labels in
$\mathbf{X}_{k-1:k}$ can be partitioned into labels terminated at\ $k-1$,
labels survived to $k$, and labels born at $k$, i.e. (\ref{eq:labelpartition3}%
) becomes
\begin{equation}%
%TCIMACRO{\tbigcup _{r=k-1}^{k}}%
%BeginExpansion
{\textstyle\bigcup_{r=k-1}^{k}}
%EndExpansion
\mathcal{L}(\mathbf{X}_{r}\mathbf{)}=(\mathcal{L(}\mathbf{X}_{k-1}%
)-\mathcal{L(}\mathbf{X}_{k}))\mathbf{\uplus(}\mathcal{L(}\mathbf{X}%
_{k-1})\cap\mathcal{L(}\mathbf{X}_{k}))\mathbf{\uplus(}\mathbb{B}_{k}%
\cap\mathcal{L(}\mathbf{X}_{k})). \label{eq:labelpartition3a}%
\end{equation}

When $i=j$, $\overleftarrow{\mathcal{L(}\mathbf{X}_{j})}=\emptyset$ and the
set of labels in $\mathbf{X}_{j:k}$ can be partitioned into live labels at $j$
and labels born after $j$, i.e. (\ref{eq:labelpartition2}) becomes%
\begin{equation}%
%TCIMACRO{\tbigcup _{r=j}^{k}}%
%BeginExpansion
{\textstyle\bigcup_{r=j}^{k}}
%EndExpansion
\mathcal{L}(\mathbf{X}_{r}\mathbf{)=}\mathcal{L}(\mathbf{X}_{j})\mathbf{\uplus
}\overrightarrow{\mathcal{L}(\mathbf{X}_{j}\mathbf{).}}
\label{eq:labelpartition4}%
\end{equation}

The following Lemma summarizes some useful properties of multi-scan exponentials.

\textbf{Lemma A.1}: Let $\mathbf{X}_{j:k}$ be a sequence of multi-object
states (generated by a set of trajectories) and $g$, $h$ be two functions
taking trajectories to the reals. Then:

(i) $\left[  gh\right]  ^{\mathbf{X}_{j:k}}=\left[  g\right]  ^{\mathbf{X}%
_{j:k}}\left[  h\right]  ^{\mathbf{X}_{j:k}}$

(ii) For a multi-object state sequence $\mathbf{Y}_{j:k}$ with labels disjoint
from those of $\mathbf{X}_{j:k}$,
\[
\left[  h\right]  ^{\mathbf{X}_{j:k}\uplus\mathbf{Y}_{j:k}}=\left[  h\right]
^{\mathbf{X}_{j:k}}\left[  h\right]  ^{\mathbf{Y}_{j:k}}%
\]

(iii) For any $i$ in $\{j:k\}$%
\[
\left[  h\right]  ^{\mathbf{X}_{j:k}}=\left[  h\right]  ^{\{\mathbf{x}%
_{s(\mathbf{\ell}):t(\mathbf{\ell})}^{(\mathbf{\ell})}:\ell\in\overleftarrow
{\mathcal{L(}\mathbf{X}_{i})}\}}\left[  h\right]  ^{\{\mathbf{x}%
_{s(\mathbf{\ell}):t(\mathbf{\ell})}^{(\mathbf{\ell})}:\ell\in\mathcal{L}%
(\mathbf{X}_{i}\mathbf{)}\}}\left[  h\right]  ^{\{\mathbf{x}_{s(\mathbf{\ell
}):t(\mathbf{\ell})}^{(\mathbf{\ell})}:\ell\in\overrightarrow{\mathcal{L}%
(\mathbf{X}_{i}\mathbf{)}}\}}%
\]
and setting $i$ to $k$, and $i$ to $j$ we have
\begin{align*}
\left[  h\right]  ^{\mathbf{X}_{j:k}}  &  =\left[  h\right]  ^{\{\mathbf{x}%
_{s(\mathbf{\ell}):t(\mathbf{\ell})}^{(\mathbf{\ell})}:\ell\in\overleftarrow
{\mathcal{L(}\mathbf{X}_{k})}\}}\left[  h\right]  ^{\{\mathbf{x}%
_{s(\mathbf{\ell}):k}^{(\mathbf{\ell})}:\ell\in\mathcal{L}(\mathbf{X}%
_{k}\mathbf{)}\},}\\
\left[  h\right]  ^{\mathbf{X}_{j:k}}  &  =\left[  h\right]  ^{\{\mathbf{x}%
_{j:t(\mathbf{\ell})}^{(\mathbf{\ell})}:\ell\in\mathcal{L}(\mathbf{X}%
_{j}\mathbf{)}\}}\left[  h\right]  ^{\{\mathbf{x}_{s(\mathbf{\ell
}):t(\mathbf{\ell})}^{(\mathbf{\ell})}:\ell\in\overrightarrow{\mathcal{L}%
(\mathbf{X}_{j}\mathbf{)}}\}}.
\end{align*}

(iv) For any $i$ in $\{j:k\}$%
\[
\left[  g\right]  ^{\mathbf{X}_{j:i}}\left[  h\right]  ^{\mathbf{X}_{i:k}%
}=\left[  g\odot h\right]  ^{\mathbf{X}_{j:k}},
\]
where%
\[
(g\odot h)(\mathbf{x}_{s(\mathbf{\ell}):t(\mathbf{\ell})}^{(\mathbf{\ell}%
)})=\left\{
\begin{array}
[c]{ll}%
h(\mathbf{x}_{s(\mathbf{\ell}):t(\mathbf{\ell})}^{(\mathbf{\ell})}) &
s(\ell)>i\\
g(\mathbf{x}_{s(\mathbf{\ell}):i}^{(\mathbf{\ell})})h(\mathbf{x}%
_{i:t(\mathbf{\ell})}^{(\mathbf{\ell})}) & s(\ell)\leq i\leq t(\ell)\\
g(\mathbf{x}_{s(\mathbf{\ell}):t(\mathbf{\ell})}^{(\mathbf{\ell})}) &
t(\ell)<i
\end{array}
\right.
\]

\textbf{Proof:} (i) and (ii) follows straight from the definition of
multi-scan exponential.

To prove (iii) noting from (\ref{eq:labelpartition0}) that the set of labels
$\cup_{r=j}^{k}\mathcal{L}(\mathbf{X}_{r}\mathbf{)}$ can be partitioned into
those terminated before $i$, live at $i$, and born after $i$, we partition the
set $\mathbf{X}_{j:k}$ of trajectories accordingly, i.e.%
\begin{align*}
\mathbf{X}_{j:k}  &  =\{\mathbf{x}_{s(\mathbf{\ell}):t(\mathbf{\ell}%
)}^{(\mathbf{\ell})}:\ell\in\cup_{r=j}^{k}\mathcal{L}(\mathbf{X}_{r}%
\mathbf{)}\}\\
&  =\{\mathbf{x}_{s(\mathbf{\ell}):t(\mathbf{\ell})}^{(\mathbf{\ell})}:\ell
\in\overleftarrow{\mathcal{L(}\mathbf{X}_{i})}\}\mathbf{\uplus}\{\mathbf{x}%
_{s(\mathbf{\ell}):t(\mathbf{\ell})}^{(\mathbf{\ell})}:\ell\in\mathcal{L(}%
\mathbf{X}_{i})\}\mathbf{\uplus}\{\mathbf{x}_{s(\mathbf{\ell}):t(\mathbf{\ell
})}^{(\mathbf{\ell})}:\ell\in\overrightarrow{\mathcal{L}(\mathbf{X}%
_{i}\mathbf{)}}\}.
\end{align*}
Hence, using (ii) gives%
\[
\left[  h\right]  ^{\mathbf{X}_{j:k}}=\left[  h\right]  ^{\{\mathbf{x}%
_{s(\mathbf{\ell}):t(\mathbf{\ell})}^{(\mathbf{\ell})}:\ell\in\overleftarrow
{\mathcal{L(}\mathbf{X}_{i})}\}}\left[  h\right]  ^{\{\mathbf{x}%
_{s(\mathbf{\ell}):t(\mathbf{\ell})}^{(\mathbf{\ell})}:\ell\in\mathcal{L}%
(\mathbf{X}_{i}\mathbf{)}\}}\left[  h\right]  ^{\{\mathbf{x}_{s(\mathbf{\ell
}):t(\mathbf{\ell})}^{(\mathbf{\ell})}:\ell\in\overrightarrow{\mathcal{L}%
(\mathbf{X}_{i}\mathbf{)}}\}}.
\]

(iv) Using the corollaries to part (iii), we partition $\left[  g\right]
^{\mathbf{X}_{j:i}}$ and $\left[  h\right]  ^{\mathbf{X}_{i:k}}$, and then
combine them as follows%
\begin{align*}
\left[  g\right]  ^{\mathbf{X}_{j:i}}\left[  h\right]  ^{\mathbf{X}_{i:k}}  &
=\left[  g\right]  ^{\{\mathbf{x}_{s(\mathbf{\ell}):t(\mathbf{\ell}%
)}^{(\mathbf{\ell})}:\ell\in\overleftarrow{\mathcal{L(}\mathbf{X}_{i})}%
\}}\left[  g\right]  ^{\{\mathbf{x}_{s(\mathbf{\ell}):i}^{(\mathbf{\ell}%
)}:\ell\in\mathcal{L(}\mathbf{X}_{i})\}}\left[  h\right]  ^{\{\mathbf{x}%
_{i:t(\mathbf{\ell})}^{(\mathbf{\ell})}:\ell\in\mathcal{L(}\mathbf{X}_{i}%
)\}}\left[  h\right]  ^{\{\mathbf{x}_{s(\mathbf{\ell}):t(\mathbf{\ell}%
)}^{(\mathbf{\ell})}:\ell\in\overrightarrow{\mathcal{L}(\mathbf{X}%
_{i}\mathbf{)}}\}}\\
&  =\left[  g\right]  ^{\{\mathbf{x}_{s(\mathbf{\ell}):t(\mathbf{\ell}%
)}^{(\mathbf{\ell})}:\ell\in\overleftarrow{\mathcal{L(}\mathbf{X}_{i})}%
\}}\prod\limits_{\ell\in\mathcal{L(}\mathbf{X}_{i})}g(\mathbf{x}%
_{s(\mathbf{\ell}):i}^{(\mathbf{\ell})})\prod\limits_{\ell\in\mathcal{L(}%
\mathbf{X}_{i})}h(\mathbf{x}_{i:t(\mathbf{\ell})}^{(\mathbf{\ell})})\left[
h\right]  ^{\{\mathbf{x}_{s(\mathbf{\ell}):t(\mathbf{\ell})}^{(\mathbf{\ell}%
)}:\ell\in\overrightarrow{\mathcal{L}(\mathbf{X}_{i}\mathbf{)}}\}}\\
&  =\left[  g\odot h\right]  ^{\{\mathbf{x}_{s(\mathbf{\ell}):t(\mathbf{\ell
})}^{(\mathbf{\ell})}:\ell\in\overleftarrow{\mathcal{L(}\mathbf{X}_{i})}%
\}}\left[  g\odot h\right]  ^{\{\mathbf{x}_{s(\mathbf{\ell}):t(\mathbf{\ell}%
)}^{(\mathbf{\ell})}:\ell\in\mathcal{L(}\mathbf{X}_{i})\}}\left[  g\odot
h\right]  ^{\{\mathbf{x}_{s(\mathbf{\ell}):t(\mathbf{\ell})}^{(\mathbf{\ell}%
)}:\ell\in\overrightarrow{\mathcal{L}(\mathbf{X}_{i}\mathbf{)}}\}}\\
&  =\left[  g\odot h\right]  ^{\mathbf{X}_{j:k}}%
\end{align*}

\subsection{Alternative form of the GLMB recursion$\label{sec:AppendixAltGLMB}%
$}

The GLMB filtering density closest to the desired form (\ref{eq:GLMB_joint0})
is given by equation (14) of \cite{VVH17}%
\begin{equation}
\mathbf{\pi}_{k\!}(\mathbf{X}_{k})\!\propto\Delta_{\!}(_{\!}\mathbf{X}%
_{k\!})\!\!\!\!\sum\limits_{I_{k-1}\!,\xi,I_{\!k},\theta_{\!k\!}%
}\!\!\!\!w^{(\xi)}(I_{k-1})\omega_{Z_{_{\!}k}}^{(_{\!}I_{k-1\!},\xi
,I_{\!k\!},\theta_{\!k\!})}\delta_{_{\!}I_{k\!}}[\mathcal{L}(_{\!}%
\mathbf{X}_{k\!})]\!\left[  p_{k_{_{\!}}}^{(_{\!}\xi,\theta_{\!k\!})}{}%
_{\!}\right]  ^{\!\mathbf{X}_{k}}\!\!\!\label{eq:AlternateGLMB}%
\end{equation}
with the weight increment given by eqs. (22), (23) of \cite{VVH17}:%
\begin{equation}
\omega_{Z_{k}}^{(I_{k-1},\xi,I_{k},\theta_{k})}=1_{{\Gamma}}(\gamma
)\prod\limits_{i=1}^{P}\eta_{i}(\gamma_{i}),\label{eq:eta_prod}%
\end{equation}
where ${\Gamma}$ is the set of positive 1-1 $P$-tuples in $\{-1$:$M\}^{P}$,
$\gamma=(\gamma_{1:P})\in\{-1$:$M\}^{P}$,%
\[
\gamma_{i}=\left\{
\begin{array}
[c]{ll}%
\theta_{k}(\ell_{i}), & \text{if }\ell_{i}\in\mathcal{D(}\theta_{k})\\
-1, & \text{otherwise}%
\end{array}
\right.  ,
\]
$\{\ell_{1:R}\}=I_{k-1}$, $\{\ell_{R+1:P}\}=\mathbb{B}$, and
\[
\eta_{i}(\gamma_{i})\!=\!%
\begin{cases}
\bar{Q}_{S,k-1}^{(\xi)}(\ell_{i}), & \!1\leq i\leq R,\text{ }\gamma
_{i}\!<0\!,\\
\bar{P}_{S,k-1\!}^{(\xi)\!}(\ell_{i})\bar{\psi}_{Z_{k_{\!}}}^{(\xi,j)\!}%
(\ell_{i\!}), & \!1\leq i\leq R,\text{ }\gamma_{i}\!\geq0,\\
Q_{B,k}(\ell_{i}), & \!R\!+\!1\leq i\leq P,\text{ }\gamma_{i}\!<0,\\
P_{B\!,k}(\ell_{i})\bar{\psi}_{Z_{k}}^{(\xi,j)}(\ell_{i}), & \!R\!+\!1\leq
i\leq P,\text{ }\gamma_{i}\!\geq0.
\end{cases}
.
\]

Further, eq. (21) of \cite{VVH17} shows that any $\gamma\in{\Gamma}$ defines a
positive 1-1 $\theta_{k}:I_{k}\rightarrow\{0$:$M\}$ by%
\[
I_{k}=\{\ell_{i}\in I_{k-1}\mathbf{\uplus}\mathbb{B}_{k}:\gamma_{i}%
\geq0\}\text{ and }\theta_{k}(\ell_{i})=\gamma_{i}.
\]
Thus, for any corresponding pair of $\gamma$ and $\theta_{k}$, we have:
\begin{align*}
\mathcal{D(}\theta_{k}) &  \triangleq I_{k}\subseteq I_{k-1}\mathbf{\uplus
}\mathbb{B}_{k};\\
1_{{\Gamma}}(\gamma) &  =1_{{\Theta}_{k}(I_{k})}(\theta_{k});\\
R\!+\!1\leq i\leq P,\text{ }\gamma_{i}\geq0 &  \Longleftrightarrow\ell_{i}%
\in\mathcal{D(}\theta_{k})\cap\mathbb{B}_{k};\\
1\leq i\leq R,\text{ }\gamma_{i}\geq0 &  \Longleftrightarrow\ell_{i}%
\in\mathcal{D(}\theta_{k})-\mathbb{B}_{k};\\
R\!+\!1\leq i\leq P,\text{ }\gamma_{i}\!<0 &  \Longleftrightarrow\ell_{i}%
\in\mathbb{B}_{k}-\mathcal{D(}\theta_{k}).
\end{align*}
Noting that $P_{B\!,k}(\ell_{i})\bar{\psi}_{Z_{k}}^{(\xi,\theta_{k}(\ell
_{i}))}(\ell_{i})=\bar{\Lambda}_{B,k}^{(\theta_{k}(\ell_{i}))}(\ell_{i})$ and
$\bar{P}_{S,k-1\!}^{(\xi)\!}(\ell_{i})\bar{\psi}_{Z_{k_{\!}}}^{(\xi,\theta
_{k}(\ell_{i}))\!}(\ell_{i\!})=\bar{\Lambda}_{S,k|k-1}^{(\xi,\theta_{k}%
(\ell_{i}))}(\ell_{i})$, (\ref{eq:eta_prod}) becomes
\[
\omega_{Z_{k}}^{(I_{k-1},\xi,I_{k},\theta_{k})}=1_{{\Theta(I_{k})}}(\theta
_{k})1_{\mathcal{F}(\mathbb{B}_{k}\mathbf{\uplus}I_{k-1})}(\mathcal{D(}%
\theta_{k}))\left[  \omega_{k|k-1}^{(\xi,\theta_{k}(\cdot))}(\cdot)\right]
^{\mathbb{B}_{k}\uplus I_{k-1}}.
\]

Further, for any $\theta_{k}\in\Theta_{k}$, $1_{\Theta_{k}(I_{k})}(\theta
_{k})\delta_{I_{k}}[\mathcal{L}(_{\!}\mathbf{X}_{k\!})]=\delta_{\mathcal{D(}%
\theta_{k})}[\mathcal{L}(_{\!}\mathbf{X}_{k\!})]$, and summing over the pair
($I_{k},\theta_{k}$) with the constraint $1_{\Theta_{k}(I_{k})}(\theta
_{k})\delta_{I_{k}}[\mathcal{L}(_{\!}\mathbf{X}_{k\!})]=1$ is the same as
summing over $\theta_{k}$ with the constraint $\delta_{\mathcal{D(}\theta
_{k})}[\mathcal{L}(_{\!}\mathbf{X}_{k\!})]=1$. Hence, (\ref{eq:AlternateGLMB})
becomes (\ref{eq:GLMB_joint0}).

\subsection{Proof of Proposition \ref{prop:transition} (Multi-object
Transition) $\label{sec:AppendixTransition}$}

Using (\ref{eq:labelpartition3a}) to partition $\mathcal{L(}\mathbf{X}%
_{k-1})\cup\mathcal{L(}\mathbf{X}_{k})$ into disappearing labels at time
$k-1$, surviving labels at time $k$, and new born labels at time $k$, and
noting that
\begin{align*}
\{\mathbf{x}_{k}^{(\mathbf{\ell})}  &  :\ell\in\mathbb{B}_{k}\cap
\mathcal{L(}\mathbf{X}_{k})\}=\{\mathbf{x}_{k-1:k}^{(\mathbf{\ell})}%
:t(\ell)=k\}\\
\{\mathbf{x}_{k-1:k}^{(\mathbf{\ell})}  &  :\ell\in\mathcal{L(}\mathbf{X}%
_{k-1})\cap\mathcal{L(}\mathbf{X}_{k})\}=\{\mathbf{x}_{k-1:k}^{(\mathbf{\ell
})}:t(\ell)=k\}\\
\{\mathbf{x}_{k-1}^{(\mathbf{\ell})}  &  :\ell\in\mathcal{L(}\mathbf{X}%
_{k-1})-\mathcal{L(}\mathbf{X}_{k})\}=\{\mathbf{x}_{k-1}^{(\mathbf{\ell}%
)}:t(\ell)=k-1\}
\end{align*}
we have%
\begin{align*}
\mathbf{X}_{k-1:k}  &  \equiv\{\mathbf{x}_{s(\mathbf{\ell}):t(\mathbf{\ell}%
)}^{(\mathbf{\ell})}:\ell\in\mathcal{L(}\mathbf{X}_{k-1})\cup\mathcal{L(}%
\mathbf{X}_{k})\}\\
&  =\{\mathbf{x}_{k-1}^{(\mathbf{\ell})}:t(\ell)=k-1\}\mathbf{\uplus
}\{\mathbf{x}_{k-1:k}^{(\mathbf{\ell})}:t(\ell)=k\}\mathbf{\uplus}%
\{\mathbf{x}_{k}^{(\mathbf{\ell})}:s(\ell)=k\}.
\end{align*}

Let $\mathbf{x}_{k}^{(\ell)}=(x_{k}^{(\ell)},\ell)$ to denote the element of
the multi-object state $\mathbf{X}_{k}$ at time $k$, with label $\ell
\in\mathcal{L(}\mathbf{X}_{k})$, then the multi-object transition density
given in \cite{VoGLMB13}, \cite{VVP_GLMB13} can be rewritten as
\begin{align*}
\mathbf{f}_{k|k-1}\left(  \mathbf{\mathbf{X}}_{k}|\mathbf{X}_{k-1}\right)   &
=\Delta(\mathbf{\mathbf{X}}_{k})1_{\mathcal{F}(\mathbb{B}_{k}\mathbf{\uplus
}\mathcal{L(}\mathbf{X}_{k-1}))}(\mathcal{L(}\mathbf{X}_{k}))Q_{B,k}%
^{\mathbb{B}_{k}-\mathcal{L(}\mathbf{\mathbf{X}}_{k})}\!\!\prod\limits_{\ell
\in\mathcal{L(}\mathbf{X}_{k-1})-\mathcal{L(}\mathbf{X}_{k})}\!\!\!Q_{S,k-1}%
(x_{k-1}^{(\ell)},\ell)\\
&  \times\prod\limits_{\ell\in\mathbb{B}_{k}\cap\mathcal{L(}\mathbf{X}_{k}%
)}\!\!\!P_{B,k}(\ell)f_{B,k}(x_{k}^{(\ell)},\ell)\prod\limits_{\ell
\in\mathcal{L(}\mathbf{X}_{k-1})\cap\mathcal{L(}\mathbf{X}_{k})}%
\!\!\!P_{S,k-1}(x_{k-1}^{(\ell)},\ell)f_{S,k|k-1}(x_{k}^{(\ell)}%
|x_{k-1}^{(\ell)},\ell)\\
&  =\Delta(\mathbf{\mathbf{X}}_{k})1_{\mathcal{F}(\mathbb{B}_{k}%
\mathbf{\uplus}\mathcal{L(}\mathbf{X}_{k-1}))}(\mathcal{L(}\mathbf{X}%
_{k}))Q_{B,k}^{\mathbb{B}_{k}-\mathcal{L(}\mathbf{\mathbf{X}}_{k})}\\
&  \times\left[  \phi_{k-1:k}\right]  ^{\{\mathbf{x}_{k-1}^{(\mathbf{\ell}%
)}:t(\ell)=k-1\}}\left[  \phi_{k-1:k}\right]  ^{\{\mathbf{x}_{k-1:k}%
^{(\mathbf{\ell})}:t(\ell)=k\}}\left[  \phi_{k-1:k}\right]  ^{\{\mathbf{x}%
_{k-1:k}^{(\mathbf{\ell})}:t(\ell)=k\}}\\
&  =\Delta(\mathbf{\mathbf{X}}_{k})1_{\mathcal{F}(\mathbb{B}_{k}%
\mathbf{\uplus}\mathcal{L(}\mathbf{X}_{k-1}))}(\mathcal{L(}\mathbf{X}%
_{k}))Q_{B,k}^{\mathbb{B}_{k}-\mathcal{L(}\mathbf{\mathbf{X}}_{k})}\left[
\phi_{k-1:k}\right]  ^{\mathbf{X}_{k-1:k}}%
\end{align*}
where the last step follows from Lemma A.1 part (ii).

\subsection{Proof of Proposition \ref{prop:expectation}%
$\label{sec:AppendixExpectation}$}

Using the $\delta$-form we have%
\begin{align}
\int f(\mathcal{L(}\mathbf{\mathbf{X}}_{j:k}))\mathbf{\pi}(\mathbf{X}%
_{j:k})\delta\mathbf{X}_{j:k}  &  =\int f(\mathcal{L(}\mathbf{\mathbf{X}%
}_{j:k}))\sum_{\xi}\sum_{I_{j:k}}w^{(\xi)}(I_{j:k})\delta_{j:k}[\mathcal{L(}%
\mathbf{X}_{j:k})]\left[  p^{(\xi)}\right]  ^{\mathbf{X}_{j:k}}\delta
\mathbf{X}_{j:k}\nonumber\\
&  =\sum_{\xi}\sum_{I_{j:k}}f(I_{j:k})w^{(\xi)}(I_{j:k})\int\delta
_{j:k}[\mathcal{L(}\mathbf{X}_{j:k})]\left[  p^{(\xi)}\right]  ^{\mathbf{X}%
_{j:k}}\delta\mathbf{X}_{j:k}\nonumber\\
&  =\sum_{\xi}\sum_{I_{j:k}}f(I_{j:k})w^{(\xi)}(I_{j:k})\prod\limits_{\ell
\in\cup_{i=j}^{k}I_{i}}\int p^{(\xi)}(x_{s(\mathbf{\ell}):t(\mathbf{\ell}%
)}^{(\ell)};\ell)dx_{s(\mathbf{\ell}):t(\mathbf{\ell})}^{(\ell)}%
\label{eq:proof-integral}\\
&  =\sum_{\xi}\sum_{I_{j:k}}f(I_{j:k})w^{(\xi)}(I_{j:k}).\nonumber
\end{align}
where (\ref{eq:proof-integral}) follows from Lemma A.2 below.

\textbf{Lemma A.2}: For a function $h$ taking trajectories to the reals, with
$h(\cdot;\ell)$ integrable for each $\ell\in\cup_{i=j}^{k}I_{i}$
\begin{equation}
\int\delta_{I_{j:k}}[\mathcal{L(}\mathbf{X}_{j:k})][h]^{\mathbf{X}_{j:k}%
}\delta\mathbf{X}_{j:k}=\prod\limits_{\ell\in\cup_{i=j}^{k}I_{i}}\int
h(x_{s(\mathbf{\ell}):t(\mathbf{\ell})}^{(\ell)};\ell)dx_{s(\mathbf{\ell
}):t(\mathbf{\ell})}^{(\ell)}. \label{eq:LemmaA20}%
\end{equation}

Proof: For $g:\mathbf{\mathbf{\mathcal{F}(}\mathbb{X}\mathcal{\times
}\mathbb{L)}}\rightarrow\mathbb{R}$ and $I=\{i_{1},...,i_{|I|}\}\subseteq
\mathbf{\mathbb{L}}$,
\begin{align}
\int\delta_{I}[\mathcal{L(}\mathbf{X})]g(\mathbf{X})\delta\mathbf{X}  &
=\sum_{n=0}^{\infty}\sum_{(l_{1},...l_{n})}\frac{1}{n!}\int\delta_{I}%
[\{l_{1},...,l_{n}\}]g(\{(l_{1},x_{1}),...,(l_{n},x_{n})\})dx_{1:n}\nonumber\\
&  =\int g(\{(i_{1},x_{1}),...,(i_{|I|},x_{|I|})\})dx_{1:|I|}
\label{eq:LemmaA2_1}%
\end{align}

For $g:\mathbf{\mathbf{\mathcal{F}(}\mathbb{X}\mathcal{\times}\mathbb{L}}%
_{j}\mathbf{\mathbb{)}\mathcal{\times}}\cdots\mathbf{\mathcal{\times F}%
(}\mathbb{X}\mathcal{\times}\mathbb{L}_{k}\mathbb{)}\rightarrow\mathbb{R}$ ,
and $I_{t}=\{i_{t,1},...,i_{t,|I_{t}|}\}\subseteq\mathbf{\mathbb{L}}%
_{t},t=j,...,k$,
\begin{align*}
\int\delta_{I_{j:k}}[\mathcal{L(}\mathbf{X}_{j:k})]g(\mathbf{X}_{j:k}%
)\delta\mathbf{X}_{j}  &  =\delta_{I_{j+1:k}}[\mathcal{L(}\mathbf{X}%
_{j+1:k})]\int\delta_{I_{j}}[\mathcal{L(}\mathbf{X}_{j})]g(\mathbf{X}%
_{j},\mathbf{X}_{j+1:k})\delta\mathbf{X}_{j}\\
&  =\delta_{I_{j+1:k}}[\mathcal{L(}\mathbf{X}_{j+1:k})]\int g(\{(i_{j,1}%
,x_{j,1}),...,(i_{j,|I_{j}|},x_{j,|I_{j}|})\},\mathbf{X}_{j+1:k}%
)dx_{j,1:|I_{j}|}%
\end{align*}
where the last line follows from (\ref{eq:LemmaA2_1}). Further, iterating for
$j+1$, ..., $k$%
\begin{align*}
&  \int\delta_{I_{j:k}}[\mathcal{L(}\mathbf{X}_{j:k})]g(\mathbf{X}%
_{j:k})\delta\mathbf{X}_{j:k}\\
&  =\int...\int g(\{(i_{j,1},x_{j,1}),...,(i_{j,N_{j}},x_{j,|I_{j}%
|})\},...,\{(i_{k,1},x_{k,1}),...,(i_{k,N_{k}},x_{k,|I_{k}|})\})dx_{j,1:|I_{j}%
|}\cdots dx_{k,1:|I_{k}|}%
\end{align*}
Setting $g(\mathbf{X}_{j:k})=[h]^{\mathbf{X}_{j:k}}$ yields%
\begin{align*}
\int\delta_{I_{j:k}}[\mathcal{L(}\mathbf{X}_{j:k})][h]^{\mathbf{X}_{j:k}%
}\delta\mathbf{X}_{j:k}  &  =\int\delta_{I_{j:k}}[\mathcal{L(}\mathbf{X}%
_{j:k})]\prod\limits_{\ell\in\cup_{i=j}^{k}I_{i}}h(\mathbf{x}_{s(\mathbf{\ell
}):t(\mathbf{\ell})}^{(\mathbf{\ell})})\delta\mathbf{X}_{j:k}\\
&  =\int...\int\prod\limits_{\ell\in\cup_{i=j}^{k}I_{i}}h(\mathbf{x}%
_{s(\mathbf{\ell}):t(\mathbf{\ell})}^{(\mathbf{\ell})})dx_{j,1:|I_{j}|}\cdots
dx_{k,1:|I_{k}|}\\
&  =\int...\int\prod\limits_{\ell\in\cup_{i=j}^{k}I_{i}}h(x_{s(\mathbf{\ell
}):t(\mathbf{\ell})}^{(\ell)};\ell)dx_{j,1}\cdots dx_{j,|I_{j}|}\cdots
dx_{k,1}\cdots dx_{k,|I_{k}|}\\
&  =\prod\limits_{\ell\in\cup_{i=j}^{k}I_{i}}\int h(x_{s(\mathbf{\ell
}):t(\mathbf{\ell})}^{(\ell)};\ell)dx_{s(\mathbf{\ell}):t(\mathbf{\ell}%
)}^{(\ell)}%
\end{align*}
where the last step follows from regrouping $dx_{j,1}\cdots dx_{j,|I_{j}%
|}\cdots dx_{k,1}\cdots dx_{k,|I_{k}|}$ to $\prod\limits_{\ell\in\cup
_{i=j}^{k}I_{i}}dx_{s(\mathbf{\ell}):t(\mathbf{\ell})}^{(\ell)}$.

\subsection{Proof of Proposition \ref{prop:MSGLMBrecursion} (Multi-Scan GLMB
Recursion) $\label{sec:AppendixMultiScanRec}$}

Substituting (\ref{eq:LemmaA.4_1}) from Lemma A.3 into the posterior
recursion:
\begin{align*}
&  \mathbf{\pi}_{0:k}(\mathbf{X}_{0:k})\propto g_{k}(Z_{k}|\mathbf{X}%
_{k})\mathbf{f}_{k|k-1}(\mathbf{X}_{k}|\mathbf{X}_{k-1})\mathbf{\pi}%
_{0:k-1}(\mathbf{X}_{0:k-1})\\
\! &  =\Delta(\mathbf{\mathbf{X}}_{0:k})\sum_{\xi,\theta_{k}}w_{0:k\!-\!1}%
^{(\xi)}(\mathcal{L(}\mathbf{X}_{0:k\!-\!1}))1_{\!\mathcal{F}(\mathbb{B}%
_{k}\mathbf{\uplus}\mathcal{L(}\mathbf{X}_{k\!-\!1}))}(\mathcal{D(}\theta
_{k}))\delta_{\mathcal{D(}\theta_{k})}[\mathcal{L(}\mathbf{X}_{k}%
)]Q_{B,k}^{\mathbb{B}_{k}-\mathcal{D(}\theta_{k})}\!\left[  p^{(\xi
)\!}\right]  ^{\!\mathbf{X}_{0:k-1}}\!\left[  \lambda_{k-1:k\!}^{(\theta_{k}%
)}\right]  ^{\!\mathbf{X}_{k-1:k}}\\
\! &  =\Delta(\mathbf{\mathbf{X}}_{0:k})\sum_{\xi,\theta_{k}}1_{\mathcal{F}%
(\mathbb{B}_{k}\mathbf{\uplus}\mathcal{L(}\mathbf{X}_{k-1}))}(\mathcal{D(}%
\theta_{k}))\delta_{\mathcal{D(}\theta_{k})}[\mathcal{L(}\mathbf{X}%
_{k})]w_{0:k\!-\!1}^{(\xi)}(\mathcal{L(}\mathbf{X}_{0:k\!-\!1}))Q_{B,k}%
^{\mathbb{B}_{k}-\mathcal{D(}\theta_{k})}\!\left[  p^{(\xi)}\odot
\lambda_{k-1:k}^{(\theta_{k})}\right]  ^{\!\mathbf{X}_{0:k}}%
\end{align*}
where the last step follows from Lemma A.1 (iv). The proof is completed by
showing
\begin{equation}
Q_{B,k}^{\mathbb{B}_{k}-\mathcal{D(}\theta_{k})}\left[  p^{(\xi)}\odot
\lambda_{k-1:k}^{(\theta_{k})}\right]  ^{\mathbf{X}_{0:k}}=\left[
p_{0:k}^{(\xi,\theta_{k})}\right]  ^{\mathbf{X}_{0:k}}\left[  \omega
_{k|k-1}^{(\xi,\theta_{k})}\right]  ^{\mathbb{B}_{k}\uplus\mathcal{L(}%
\mathbf{X}_{k-1})},\label{eq:ProofofProp6}%
\end{equation}
which is simply a matter of algebra. For completeness, we substitute
(\ref{eq:LemmaA.4_2}) for $\lambda_{k-1:k}^{(\theta_{k})}(\cdot;\ell)$ into
$p^{(\xi)}\odot\lambda_{k-1:k}^{(\theta_{k})}$:
\begin{align*}
(p^{(\xi)}\odot\lambda_{k-1:k}^{(\theta_{k})})(x_{s(\ell):t(\ell)}^{(\ell
)};\ell)\! &  \triangleq\!\left\{  \!\!%
\begin{array}
[c]{ll}%
\lambda_{k-1:k}^{(\theta_{k})}(x_{k}^{(\ell)},\ell), & s(\ell)>k-1\\
p^{(\xi)}(x_{s(\ell):k-1}^{(\ell)};\ell)\lambda_{k-1:k}^{(\theta_{k}%
)}(x_{k-1:t(\ell)}^{(\ell)};\ell), & s(\ell)\leq k-1\leq t(\ell)\\
p^{(\xi)}(x_{s(\ell):t(\ell)}^{(\ell)};\ell), & t(\ell)<k-1
\end{array}
\right.  \\
&  =\!\left\{  \!\!%
\begin{array}
[c]{ll}%
\Lambda_{\!B,k}^{(\theta_{k}(\ell))}(x_{k}^{(\ell)};\ell),\text{ } &
\!s(\ell)=k\\
p^{(\xi)}(x_{s(\ell):k-1}^{(\ell)};\ell)\Lambda_{\!S,k|k-1}^{(\theta_{k}%
(\ell))}(x_{k-1}^{(\ell)},x_{k}^{(\ell)};\ell), & \!t(\ell)=k>s(\ell)\\
p^{(\xi)}(x_{s(\ell):k-1}^{(\ell)};\ell)Q_{S,k-1}(x_{k-1}^{(\ell)},\ell), &
\!t(\ell)=k-1\geq s(\ell)\\
p^{(\xi)}(x_{s(\ell):t(\ell)}^{(\ell)};\ell), & \!t(\ell)<k-1
\end{array}
\right.  \\
&  =\!p_{0:k}^{(\xi,\theta_{k})}(x_{s(\ell):t(\ell)}^{(\ell)};\ell
)\times\left\{  \!\!%
\begin{array}
[c]{ll}%
\!\bar{\Lambda}_{B,k}^{(\theta_{k}(\ell))}(\ell), & \!s(\ell)=k\\
\!\bar{\Lambda}_{S,k|k-1}^{(\xi,\theta_{k}(\ell))}(\ell), & \!t(\ell
)=k>s(\ell)\\
\!\bar{Q}_{S,k-1}^{(\xi)}(\ell), & \!t(\ell)=k-1\geq s(\ell)\\
\!1, & \!t(\ell)<k-1
\end{array}
\right.
\end{align*}
note that in the last step we used the definition of $p_{0:k}^{(\xi,\theta
_{k})}(x_{s(\ell):t(\ell)}^{(\ell)};\ell)$.

Consequently,%
\[
\left[  p^{(\xi)}\odot\lambda_{k-1:k}^{(\theta_{k})}\right]  ^{\mathbf{X}%
_{0:k}}=\left[  p_{0:k}^{(\xi,\theta_{k})}\right]  ^{\mathbf{X}_{0:k}}%
\prod\limits_{\substack{\ell\in\cup_{r=0}^{k}\mathcal{L}(\mathbf{X}%
_{r}\mathbf{)}\\t(\ell)=k-1\geq s(\ell)}}\!\!\bar{Q}_{S,k-1}^{(\xi)}%
(\ell)\!\!\prod\limits_{\substack{\ell\in\cup_{r=0}^{k}\mathcal{L}%
(\mathbf{X}_{r}\mathbf{)}\\t(\ell)=k>s(\ell)}}\!\!\bar{\Lambda}_{S,k|k-1}%
^{(\xi,\theta_{k}(\ell))}(\ell)\!\!\prod\limits_{\substack{\ell\in\cup
_{r=0}^{k}\mathcal{L}(\mathbf{X}_{r}\mathbf{)}\\s(\ell)=k}}\!\!\bar{\Lambda
}_{B,k}^{(\theta_{k}(\ell))}(\ell)
\]
Multiplying by $Q_{B,k}^{\mathbb{B}_{k}-\mathcal{D(}\theta_{k})}$ and using
the following equivalences: $t(\ell)=k-1\geq s(\ell)$ iff $\ell\in
\mathcal{L}(\mathbf{X}_{k-1}\mathbf{)-}\mathcal{D(}\theta_{k})$;
$t(\ell)=k>s(\ell)$ iff $\ell\in\mathcal{D(}\theta_{k})-\mathbb{B}_{k}$;
$s(\ell)=k$ iff $\ell\in\mathcal{D(}\theta_{k})\cap\mathbb{B}_{k}$, we have%
\begin{align*}
&  Q_{B,k}^{\mathbb{B}_{k}-\mathcal{D(}\theta_{k})}\!\left[  p^{(\xi)\!}%
\odot\!\lambda_{k-1:k}^{(\theta_{k})}\right]  ^{\mathbf{X}_{0:k}}\\
&  =\left[  p_{0:k}^{(\xi,\theta_{k})}\right]  ^{\mathbf{X}_{0:k}}%
\!\!\prod\limits_{\ell\in\mathcal{L}(\mathbf{X}_{k\!-\!1}\mathbf{)-}%
\!\mathcal{D(}\theta_{k})}\!\!\!\!\bar{Q}_{S,k-1}^{(\xi)}(\ell)\!\!\prod
\limits_{\ell\in\mathcal{D(}\theta_{k})-\mathbb{B}_{k}}\!\!\!\!\bar{\Lambda
}_{S,k|k-1}^{(\xi,\theta_{k}(\ell))}(\ell)\!\!\prod\limits_{\ell
\in\mathcal{D(}\theta_{k})\cap\mathbb{B}_{k}}\!\!\!\!\bar{\Lambda}%
_{B,k}^{(\theta_{k}(\ell))}(\ell)\!\!\prod\limits_{\ell\in\mathbb{B}%
_{k}-\mathcal{D(}\theta_{k})}\!\!\!\!Q_{B,k}(\ell)\\
&  =\left[  p_{0:k}^{(\xi,\theta_{k})}\right]  ^{\mathbf{X}_{0:k}}\left[
\omega_{k|k-1}^{(\xi,\theta_{k})}\right]  ^{\mathbb{B}_{k}\uplus
\mathcal{L(}\mathbf{X}_{k-1})}%
\end{align*}
since $\mathcal{L}(\mathbf{X}_{k-1}\mathbf{)-}\mathcal{D(}\theta_{k})$,
$\mathcal{D(}\theta_{k})-\mathbb{B}_{k}$, $\mathcal{D(}\theta_{k}%
)\cap\mathbb{B}_{k}$,\ and $\mathbb{B}_{k}-\mathcal{D(}\theta_{k})$\ form a
partition of $\mathbb{B}_{k}\uplus\mathcal{L(}\mathbf{X}_{k-1})$.

\textbf{Lemma A.3}:
\begin{equation}
g_{k}(Z_{k}|\mathbf{X}_{k})\mathbf{f}_{k|k\!-\!1}(\mathbf{X}_{k}%
|\mathbf{X}_{k\!-\!1})=\Delta(\mathbf{\mathbf{X}}_{k})\!\!\sum_{\theta_{k}%
\in\Theta_{k}}\!\!1_{\mathcal{F}(\mathbb{B}_{k}\mathbf{\uplus}\mathcal{L(}%
\mathbf{X}_{k\!-\!1}))}(\mathcal{D(}\theta_{k}))\delta_{\mathcal{D(}\theta
_{k})}[\mathcal{L(}\mathbf{X}_{k})]Q_{B,k}^{\mathbb{B}_{k}\!-\!\mathcal{D(}%
\theta_{k})}\!\left[  \lambda_{k\!-\!1:k}^{(\theta_{k})}\right]
^{\mathbf{X}_{k\!-\!1:k}} \label{eq:LemmaA.4_1}%
\end{equation}
where%
\begin{equation}
\lambda_{k-1:k}^{(\theta_{k})}(x_{s(\ell):t(\ell)}^{(\ell)};\ell)=\left\{
\begin{array}
[c]{ll}%
\Lambda_{\!B,k}^{(\theta_{k}(\ell))}(x_{k}^{(\ell)};\ell), & s(\ell)=k\\
\Lambda_{\!S,k|k-1}^{(\theta_{k}(\ell))}(x_{k-1}^{(\ell)},x_{k}^{(\ell)}%
;\ell) & t(\ell)=k>s(\ell)\\
Q_{S,k-1}(x_{k-1}^{(\ell)},\ell), & t(\ell)=k-1
\end{array}
\right.  \label{eq:LemmaA.4_2}%
\end{equation}

\textbf{Proof: }Noting that for any $\theta_{k}\in\Theta_{k}$, $1_{\Theta
_{k}(\mathcal{L(}\mathbf{X}_{k}))}(\theta_{k})=\delta_{\mathcal{D(}\theta
_{k})}[\mathcal{L}(_{\!}\mathbf{X}_{k\!})]$ we have%
\begin{align*}
&  g_{k}(Z_{k}|\mathbf{X}_{k})\mathbf{f}_{k|k-1}(\mathbf{X}_{k}|\mathbf{X}%
_{k-1})\\
&  =\Delta(\mathbf{\mathbf{X}}_{k})\sum_{\theta_{k}\in\Theta_{k}%
}1_{\mathcal{F}(\mathbb{B}_{k}\mathbf{\uplus}\mathcal{L(}\mathbf{X}_{k-1}%
))}(\mathcal{L(}\mathbf{X}_{k}))\delta_{\mathcal{D(}\theta_{k})}%
[\mathcal{L}(_{\!}\mathbf{X}_{k\!})]Q_{B,k}^{\mathbb{B}_{k}-\mathcal{L(}%
\mathbf{\mathbf{X}}_{k})}\left[  \phi_{k-1:k}\right]  ^{\mathbf{X}_{k-1:k}%
}\left[  \psi_{k,Z_{k}}^{(\theta_{k}\circ\mathcal{L}(\cdot))}(\cdot)\right]
^{\mathbf{X}_{k}}\\
&  =\Delta(\mathbf{\mathbf{X}}_{k})\sum_{\theta_{k}\in\Theta_{k}%
}1_{\mathcal{F}(\mathbb{B}_{k}\mathbf{\uplus}\mathcal{L(}\mathbf{X}_{k-1}%
))}(\mathcal{D(}\theta_{k}))Q_{B,k}^{\mathbb{B}_{k}-\mathcal{D(}\theta_{k}%
)}\delta_{\mathcal{D(}\theta_{k})}[\mathcal{L}(_{\!}\mathbf{X}_{k\!})]\left[
\phi_{k-1:k}\odot\psi_{k,Z_{k}}^{(\theta_{k}\circ\mathcal{L})}\right]
^{\mathbf{X}_{k-1:k}}%
\end{align*}
where the last step follows from Lemma A.1 (iv). The proof is completed by
showing $\phi_{k-1:k}\odot\psi_{k,Z_{k}}^{(\theta_{k}\circ\mathcal{L}%
)}=\lambda_{k-1:k}^{(\theta_{k})}$, which is simply a matter of algebra.
Substituting (\ref{eq:labeled_transition_MS1}) for $\phi_{k-1:k}$:
\begin{align*}
(\phi_{k-1:k}\odot\psi_{k,Z_{k}}^{(\theta_{k}\circ\mathcal{L})})(x_{s(\ell
):t(\ell)}^{(\ell)};\ell)\! &  \triangleq\!\left\{  \!\!%
\begin{array}
[c]{ll}%
\phi_{k-1:k}(x_{s(\ell):k}^{(\ell)};\ell)\psi_{k,Z_{k}}^{(\theta(\ell))}%
(x_{k}^{(\ell)},\ell), & \!s(\ell)\leq k\leq t(\ell)\\
\phi_{k-1:k}(x_{k-1}^{(\ell)},\ell), & \!t(\ell)=k-1
\end{array}
\right.  \\
&  =\!\left\{  \!\!%
\begin{array}
[c]{ll}%
P_{B,k}(\ell)f_{B,k}(x_{k}^{(\ell)},\ell)\psi_{k,Z_{k}}^{(\theta_{k}(\ell
))}(x_{k}^{(\ell)},\ell), & \!\!s(\ell)=k\\
P_{S,k-1}(x_{k-1}^{(\ell)},\ell)f_{S,k|k-1}(x_{k}^{(\ell)}|x_{k-1}^{(\ell
)},\ell)\psi_{k,Z_{k}}^{(\theta_{k}(\ell))}(x_{k}^{(\ell)},\ell), &
\!\!t(\ell)=k>s(\ell)\\
Q_{S,k-1}(x_{k-1}^{(\ell)},\ell), & \!\!t(\ell)=k-1
\end{array}
\right.  \\
&  =\!\left\{  \!\!%
\begin{array}
[c]{ll}%
\Lambda_{\!B,k}^{(\theta_{k}(\ell))}(x_{k}^{(\ell)};\ell), & s(\ell)=k\\
\Lambda_{\!S,k|k-1}^{(\theta_{k}(\ell))}(x_{k-1}^{(\ell)},x_{k}^{(\ell)}%
;\ell) & t(\ell)=k>s(\ell)\\
Q_{S,k-1}(x_{k-1}^{(\ell)},\ell), & t(\ell)=k-1
\end{array}
\right.  \\
&  =\!\lambda_{k-1:k}^{(\theta_{k})}(x_{s(\ell):t(\ell)}^{(\ell)};\ell).
\end{align*}

\subsection{Proof of Proposition \ref{conditionals}
$\label{sec:AppendixConditionals}$}

We note the following conditions.

\textbf{Lemma A.4.} If $\gamma_{j}$, $j\in\{1:k\}$, is an association map of a
valid association history $\gamma_{1:k}$, then%
\begin{gather}
\forall\ell\in\mathbb{L}_{j}-\mathbb{B}_{j}\uplus\mathcal{L(}\gamma
_{j-1}),\gamma_{j}(\ell)=-1\label{eq:live couldnthave died earlier}\\
\forall\ell\in\mathbb{L}_{j},\gamma_{j}(\ell)\geq0\text{ or }\gamma
_{\min\{j+1,k\}}(\ell)=-1, \label{eq:dead cant relive later}%
\end{gather}
violation of either of these conditions results in $\pi(\gamma_{1:k})=0$.

\textbf{Proof:} If there exist an $\ell\in\mathbb{L}_{j}-\mathbb{B}%
_{j}\mathbf{\uplus}\mathcal{L(}\gamma_{j-1})$ such that $\gamma_{j}(\ell
)\geq0$, then $\mathcal{L}(\gamma_{j})$ is not in $\mathbb{B}_{j}%
\uplus\mathcal{L(}\gamma_{j-1})$, i.e. $1_{\mathcal{F}(\mathbb{B}%
_{j}\mathbf{\uplus}\mathcal{L(}\gamma_{j-1}))}(\mathcal{L}(\gamma_{j}))=0$ and
so $\pi(\gamma_{1:k})=0$. Hence $\pi(\gamma_{1:k})>0$ implies
(\ref{eq:live couldnthave died earlier}).

Except for $j=k$, if there exist an $\ell\in\mathbb{L}_{j}$ such that
$\gamma_{j}(\ell)<0$ and $\gamma_{j+1}(\ell)\geq0$, then $\ell$ is not in
$\mathcal{L}(\gamma_{j})$ and $\mathcal{L}(\gamma_{j+1})$ (which contains
$\ell$) is not contained in $\mathbb{B}_{j+1}\uplus\mathcal{L(}\gamma_{j})$,
i.e. $1_{\mathcal{F}(\mathbb{B}_{j+1}\uplus\mathcal{L(}\gamma_{j}%
))}(\mathcal{L}(\gamma_{j+1}))=0$, and consequently $\pi(\gamma_{1:k})=0$.
Hence $\pi(\gamma_{1:k})>0$ implies (\ref{eq:dead cant relive later}).

To prove Proposition \ref{conditionals}, we  derive the conditional
probability
\[
\pi(\gamma_{j}(\ell_{n})|\gamma_{j}(\ell_{\bar{n}}),\gamma_{\bar{j}}%
)=\frac{\pi(\gamma_{j}(\ell_{n}),\gamma_{j}(\ell_{\bar{n}}),\gamma_{\bar{j}}%
)}{\sum_{\alpha}\pi(\gamma_{j}(\ell_{n})=\alpha,\gamma_{j}(\ell_{\bar{n}%
}),\gamma_{\bar{j}})}%
\]
for $\ell_{n}\in\{\ell_{1},...,\ell_{|\mathbb{B}_{j}\uplus\mathcal{L(}%
\gamma_{j-1})|}\}$ first, and subsequently for $\ell_{n}$ $\in$ $\{\ell
_{|\mathbb{B}_{j}\uplus\mathcal{L(}\gamma_{j-1})|+1},...,\ell_{|\mathbb{L}%
_{j}|}\}$, if this set is non-empty.

For any $\ell_{n}\in\{\ell_{1},...,\ell_{|\mathbb{B}_{j}\uplus\mathcal{L(}%
\gamma_{j-1})|}\}$, either: (i) $\gamma_{j}(\ell_{n})<0$ and $\gamma
_{\min\{j+1,k\}}(\ell_{n})\geq0$; or (ii) $\gamma_{j}(\ell_{n})\geq0$ or
$\gamma_{\min\{j+1,k\}}(\ell_{n})=-1$. However, case (i) is not possible for a
valid $\gamma_{1:k}$ because it violates (\ref{eq:dead cant relive later}),
and in turn, the validity of $\gamma_{1:k}$ (see Lemma A.5). Hence $\pi
(\gamma_{j}(\ell_{n})|\gamma_{j}(\ell_{\bar{n}}),\gamma_{\bar{j}})$ must be 0.

For case (ii), using (\ref{eq:MSGLMBCannonical1}) for the joint distribution
of valid $\gamma_{1:k}$, we have
\begin{align*}
\pi(\gamma_{j}(\ell_{n})|\gamma_{j}(\ell_{\bar{n}}),\gamma_{\bar{j}})  &
\propto1_{\Gamma_{j}}(\gamma_{j})\prod\limits_{i=j}^{k}\left[  \omega
_{i|i-1}^{(\gamma_{0:i}(\cdot))}\right]  ^{\mathbb{B}_{i}\uplus\mathcal{L(}%
\gamma_{i-1})}\\
&  \propto\left\{
\begin{array}
[c]{ll}%
\prod\limits_{i=j}^{k}\omega_{i|i-1}^{(\gamma_{0:i}(\ell_{n}))} & \gamma
_{j}(\ell_{n})\leq0\\
\prod\limits_{i=j}^{k}\omega_{i|i-1}^{(\gamma_{0:i}(\ell_{n}))}(1-1_{\gamma
_{j}(\ell_{\bar{n}})}(\gamma_{j}(\ell_{n}))) & \gamma_{j}(\ell_{n})>0
\end{array}
\right.  ,
\end{align*}
where the last step invokes Proposition 3 of \cite{VVH17}. Decomposing
$\gamma_{j}(\ell_{n})\leq0$ into two cases $\gamma_{j}(\ell_{n})=0$ and
$\gamma_{j}(\ell_{n})<0$, and combining the latter with case (i) we have
(\ref{eq:FullGibbsCond}).

For any $\ell_{n}$ $\in$ $\{\ell_{|\mathbb{B}_{j}\uplus\mathcal{L(}%
\gamma_{j-1})|+1},...,\ell_{|\mathbb{L}_{j}|}\}$, the validity of
$\gamma_{1:k}$ implies that $\gamma_{j}(\ell_{n})=-1$ and $\gamma
_{\min\{j+1,k\}}(\ell_{n})=-1$ since any other values for $\gamma_{j}(\ell
_{n})$ and $\gamma_{\min\{j+1,k\}}(\ell_{n})$ would violate either
(\ref{eq:dead cant relive later}) or (\ref{eq:live couldnthave died earlier}),
and in turn, the validity of $\gamma_{1:k}$ (see Lemma A.5). Hence we have
(\ref{eq:FullGibbsCond1}).

\end{document}